# Friendship and Natural Selection

Nicholas A. Christakis[1,2] and James H. Fowler[3,4*]

[1] Department of Medicine, Yale University, New Haven, CT 06520, USA
[2] Department of Sociology, Yale University, New Haven, CT 06520, USA
[3] Medical Genetics Division, University of California, San Diego, CA 92103, USA
[4] Political Science Department, University of California, San Diego, CA 92103, USA

[*] To whom correspondence should be addressed. e-mail: fowler@ucsd.edu

**More than any other species, humans form social ties to individuals who are neither kin nor mates, and these ties tend to be with similar people. Here, we show that this similarity extends to genotypes. Across the whole genome, friends' genotypes at the SNP level tend to be positively correlated (homophilic); however, certain genotypes are negatively correlated (heterophilic). A focused gene set analysis suggests that some of the overall correlation can be explained by specific systems; for example, an olfactory gene set is homophilic and an immune system gene set is heterophilic. Finally, homophilic genotypes exhibit significantly higher measures of positive selection, suggesting that, on average, they may yield a synergistic fitness advantage that has been helping to drive recent human evolution.**

Human social interactions, and the networks they give rise to, show striking structural regularities (*1,2*), even when comparing modernized networks to those in hunter-gatherer societies (*3*). Indeed, friendship is a fundamental characteristic of human beings (*3,4,5*), and genes are known to play a role in the formation (*6*), attributes (*7*), and structures (*8*) of friendship ties. Social ties also evince *homophily*, the tendency of people to form connections with phenotypically similar others (*9*). Evolutionary models suggest that homophily can evolve under a wide range of conditions if there is a fitness advantage to same-type interactions (*10,11*). And candidate gene studies (*12,13*) have recently identified one gene variant that exhibits positive correlation or similarity between friends (homophily) and another variant that exhibits negative correlation or dissimilarity (heterophily). It remains unclear, however, whether this phenomenon extends to multiple genotypes across the whole genome, and it is not known what role genotypic correlation may have played in human evolution.

There are (at least) four possible reasons that friends may exhibit homophily in their genotypes (*12*). First, correlation in genotypes may be a trivial by-product of the tendency of people to make friends with geographically proximate or ethnoracially similar individuals who also tend to share the same ancestry. Thus, it is important to use strict controls for population stratification in tests of genetic correlation (below, we rely on the widely used principal components method to control for ancestry). Second, people may actively choose friends of a similar genotype or terminate friendships with people who have different genotypes ("birds of a feather flock together"). This may take place via a variety of mechanisms; for example, while it is unlikely that people would observe the actual genotypes of others around them, they can observe and prefer certain phenotypes, and these may obviously be influenced by specific genotypes. It is well known that people prefer to associate with others they resemble phenotypically,(*9*) but what is not known is how this translates to the single-nucleotide polymorphism (SNP) level. Third, people may actively choose particular environments, and, in those environments, they may be more likely to encounter people with similar phenotypes influenced by specific genotypes. If people then choose friends from within these environments (even at random), it would tend to generate correlated genotypes. Fourth, people may be chosen by third parties or otherwise selected into environments or circumstances where they then come into contact with similar people. These four reasons are not mutually exclusive, of course, and they may operate in parallel; two people may become friends both through active choice of each other and active or passive choice of a convivial environment.

In contrast, there are fewer reasons that friends may exhibit heterophily in their genotypes (*12*). For example, heterophily is not likely to arise by population stratification, nor by a simple process of people choosing, or being drawn to, the same environment for the same reason. Instead, there are two other processes that might be at work. First, people may actively choose to befriend people of a different type ("opposites attract"). Second, certain environments might foster interactions between individuals with dissimilar traits.

Importantly, all of these processes may be at work simultaneously, and humans may select friends and environments based on a wide variety of traits, some of which result in advantages when homophily is present (synergy) and others of which may yield advantages to heterophily (complementarity or specialization) (*3,11*). The people to whom we are connected provide important capabilities, from the ability to ward off infections, to the ability to transmit or exploit useful information, to the ability to reciprocate cooperative exchanges. Consequently, the fitness advantage of some gene variants might be influenced by their parallel presence or absence in other individuals to whom a person is connected.

Evolutionary models show that preferences for both homophily and heterophily can evolve depending on the relative fitness advantages of genotypic similarity and dissimilarity on given traits (*10*). However, these models also show that homophily evolves under a much wider variety of conditions than heterophily – even when the fitness advantage to dissimilarity exceeds the fitness advantage to similarity (*10*). The reason is that it is easier to find and successfully interact with a similar partner in a population of similar individuals than it is to find and successfully interact with a dissimilar partner in a population of dissimilar individuals. For an intuition regarding this observation, consider populations at fixation. For populations with an advantage to homophily, all individuals have the same trait at fixation, and so they all will gain the advantage in every interaction. In contrast, for populations with an advantage to heterophily, some individuals have one trait and some have another, meaning there are still likely to be at least some same-type encounters in the population that do not yield the advantage to dissimilar-type interactions. These theoretical models thus suggest that we should find more genotypes that are positively correlated between friends than negatively correlated, and that we should, on average, expect friends to exhibit greater genetic similarity across the genome as a whole (*10*).

If homophily generally contributes to evolutionary fitness across a wide variety of traits, then we would also expect to see signs of positive selection for genes that exhibit positive correlation between friends. If so, it would suggest that our capacity to make friends with unrelated strangers may have played a role in human evolution. This capacity to form friendships and this preference for homophily (which is also seen in other social animals such as dolphins (*14*) and primates (*15*)) may possibly reflect the extended workings of a kin detection system (*16*) such that genetically similar (but unrelated) friends are a kind of "functional kin." Humans may – when choosing friends from among individuals who are *strictly not related* to them – come to choose individuals who do, after all, resemble them on a genotypic level.

Here, we conduct the first genome-wide analysis of correlation in genotypes between friends. We emphasize that we are not conducting a GWAS of a propensity to be friendly (or some similar complex social trait) here; rather, we are using GWAS techniques to identify certain theorized patterns (*10*) across the whole genome. Using data from the Framingham Heart Study, we analyze 466,608 (unimputed) SNPs in 1,932 unique subjects who are in one or more of 1,367 friendship pairs (see SI for data construction and summary). The data we use (which we have uploaded to a shared data repository at http://www.ncbi.nlm.nih.gov/projects/gap/cgi-bin/study.cgi?study_id=phs000153.v6.p5 ) are exceedingly scarce; we know of no other data set of any significant size that has information on both friendship ties and common genetic variants across the whole genome (see SI). As a check against false positives, beyond the other procedures described below, we also perform a split-sample replication study. We also emphasize that, as in other whole-genome investigations with circumscribed samples (*17,18*), our interest is not in any particular SNP, but rather in the pattern across the whole genome.

To assess general, overall homophily within pairs of friends, we calculate the *kinship* coefficient (*19*) (the probability that two alleles sampled at random from two individuals are identical by state), a measure that is equal to half the relatedness measure used in genome-wide complex trait analysis (GCTA) approaches (*20*) (though keep in mind that the pairs of friends here are *not* actually related). Positive values for this measure indicate genotypes are positively correlated, perhaps because two individuals are related, and negative values indicate two individuals are not related and, in fact, tend to have opposite genotypes. To measure heterophily, we calculate the empirical probability that two individuals have opposite genotypes at a given SNP, measured by the proportion of SNPs for which neither allele is identical by state.

For comparison, we also calculate these measures for all non-kin "stranger" pairs using the same set of 1,932 subjects who are in the friends sample. After removing kin (who can, of course, be identified using genotyping) and after removing pairs who had a social relationship (i.e., friends, spouses, etc.), we identified 1,196,429 stranger pairs (see SI). Fig.1a shows that the distribution of kinship coefficients for friends is shifted right relative to the strangers. A simple difference-in-means test suggests that friends tend to be significantly more genetically "related" than strangers (+0.0014, $p < 2 \times 10^{-16}$), and, as a benchmark, the size of the difference roughly corresponds to the kinship coefficient we would expect for fourth cousins (0.0010). This difference cannot be explained by the ancestral composition of the sample or by cryptic relatedness, since the same people are used in both the friends and strangers samples (the only thing that differs is the set of relationships between them); and we emphasize again that we can be sure these pairs of friends are not, in fact, distant cousins since they are strictly unrelated and there is no identity by descent. Meanwhile, Fig.1b shows that friends also tend to have fewer SNPs where the genotypes are exactly opposite (–0.0002, $p = 4 \times 10^{-9}$). Both of these results indicate that pairs of (strictly unrelated) friends generally tend to be more genetically homophilic than pairs of strangers from the same population, but the weaker results for opposite genotypes suggest that this general tendency towards homophily may be obscuring a tendency for some specific parts of the genome to be heterophilic.

The results so far do not control for population stratification because we wanted to characterize overall similarity. However, it is important to remember that some of the similarity in genotypes can be explained by simple assortment into relationships with people who have the same ancestral background. The Framingham Heart Study is composed of mostly whites (e.g., of Italian descent), so it is possible that a simple preference for ethnically similar others could explain the results in Fig.1. However, in the following results, we apply strict controls for population stratification to ensure that any correlation we observe is not due to such a process.

To characterize the genotypes that are most likely to be homophilic or heterophilic, we conduct a genome-wide association study (GWAS) regressing subject's expected genotype on friend's expected genotype for 1,468,013 common SNPs (minor allele frequency > 0.10, see SI for imputation and regression details); for this GWAS analysis, we use both unimputed and imputed SNPs in order to improve power, but we emphasize, again, that our interest here is not in any particular SNP, but rather in the pattern across the whole genome.

Although the individuals in the Framingham Heart Study are almost all of European

ancestry, population stratification has been shown to be a concern even in samples of European Americans (*21*). Relying on a widely used procedure to control for population stratification, we calculate the first ten principal components of the subject-gene matrix with EIGENSTRAT (*22*). None of our subjects are classified as outliers, defined as individuals whose score is at least six standard deviations from the mean on one of the top ten principal components. Nonetheless, consistent with past approaches (*22*), we include all ten principal components for both the subject and the subject's friend (20 variables in all) as controls for ancestry in each regression (see SI).

To eliminate the possibility that the results are influenced by people tending to make friends with distant relatives, we use *only* the 907 friend pairs where kinship $\leq 0$ (recall that kinship can be less than zero when unrelated individuals tend to have negatively correlated genotypes). This procedure ensures that pairs of friends in the GWAS are not actually biologically related at all. It also allows us to set aside the remaining 458 pairs of friends for a split-sample replication analysis (discussed below). However, note that this procedure biases *against* finding homophilic SNPs since it means the average correlation between friends will be weakly negative.

Finally, we guard against false positives by conducting an additional "strangers" GWAS for comparison with the "friends" GWAS. For the strangers analysis, we draw 907 random pairs from the stranger sample and, to maintain comparability, we also restrict these stranger pairs to have a kinship $\leq 0$ (see SI). Importantly, both the friends GWAS and the strangers GWAS contain exactly the same people and genotypes – only the relationships between these people are different (friends vs. strangers).

Fig.2a shows QQ plots of observed versus expected *p* values for both GWAS. We would expect some variance inflation because of the restriction on the kinship coefficient to pairs that show no positive relatedness; the average correlation in genotypes resulting from this restriction is slightly negative (mean kinship = –0.003), and this causes an excess number of markers to show negative correlation and low *p* values. To establish a baseline for this effect, we first measure the variance inflation factor in the strangers GWAS ($\lambda$ = 1.020) and note in Fig.2a that there is a slight upward shift that corroborates this tendency.

In contrast, the friends GWAS is shifted even higher and yields even lower *p* values than expected for many SNPs. In fact, the variance inflation for friends is more than double, at $\lambda$ = 1.046, in spite of the fact that the two GWAS were generated using exactly the same regression model specification. This shift is what we would expect if there were widespread low-level genetic correlation in friends across the genome, and it is consistent with recent work that shows that polygenic traits can generate inflation factors of these magnitudes (*23*). As supporting evidence for this interpretation, notice that Fig.2a shows there are many more outliers for the friends group than there are for the comparison stranger group, especially for *p* values less than $10^{-4}$. This suggests that polygenic homophily and/or heterophily, rather than sample selection, population stratification, or model misspecification, account for at least some of the inflation, and hence that a relatively large number of SNPs are significantly correlated between pairs of friends (albeit each with probably small effects) across the whole genome.

To explore more fully this difference in results between the friends and strangers GWAS, in Fig.2b, we compare their *t* statistics to see whether the differences in *p* values are driven by homophily (positive correlation) or heterophily (negative correlation). The results show that the friends GWAS yields significantly more outliers than the comparison stranger group for both homophily (Kolmogorov-Smirnov test, $p = 4 \times 10^{-3}$) and heterophily ($p < 2 \times 10^{-16}$).

Although a few individual SNPs were genome-wide significant (see SI), our interest is not in individual SNPs *per se*; and the homophily present across the whole genome, coupled with the evidence that friends exhibit both more genetic homophily and heterophily than strangers, suggest that there are many genes with low levels of correlation. In fact, we can use the measures of correlation from the friends GWAS to create a "friendship score" that can be used to predict whether two people are likely to be friends in a hold-out replication sample based on the extent to which their genotypes resemble each other (see SI). This replication sample contains 458 friend pairs and 458 stranger pairs that were not used to fit the GWAS models (see SI). The results show that a one-standard-deviation change in the friendship score derived from the GWAS on the original friends sample increases the probability that a pair in the replication sample is friends by 6% ($p = 2 \times 10^{-4}$) and it can explain approximately 1.4% of the variance in the existence of friendship ties. This is similar to the variance explained using the best currently available genetic scores for schizophrenia and bipolar disorder (0.4% to 3.2%) (*24*) and body-mass index (1.5%) (*25*). Although no other large datasets with fully genotyped friends exist at this time, we expect that a future GWAS on larger samples of friends might help to improve these friendship scores, boosting both efficiency and variance explained out of sample.

We expect that there are likely to be dozens and maybe even hundreds of genetic pathways that form the basis of correlation in specific genotypes, and our sample gives us enough power to detect a few of these pathways. We first conduct a gene-based association test of the likelihood that the set of SNPs within 50 kilobase pairs of each of 17,413 genes exhibit (1) homophily or (2) heterophily (see SI). We then aggregate these results to conduct a gene-set analysis in order to determine if the most significantly homophilic and heterophilic genes are overrepresented in any functional pathways documented in the KEGG and GOSlim databases (see SI). In addition to examining the top 1% most homophilic and most heterophilic genes, we also examine the top 25% because highly polygenic traits may exhibit small differences across a large number of genes (*26*), and we expect homophily to be highly polygenic based on prior theoretical work (*10*).

Table 1 shows that three gene sets are significantly overrepresented in these analyses, after adjusting for multiple testing. In the 174 most homophilic genes (top 1%), we find that an olfactory transduction pathway is significantly overrepresented ($p = 4 \times 10^{-5}$, adjusted $p = 0.009$), suggesting that friends tend to have genotypes that yield similar senses of smell. When we increase the threshold to the top 25% of homophilic genes, we also find that the linoleic acid metabolism system is significantly overrepresented ($p = 2 \times 10^{-5}$, adjusted $p = 0.005$). For heterophilic genes, a gene set that characterizes certain immune system processes achieves significance ($p = 5 \times 10^{-4}$, adjusted $p = 0.036$), which suggests that friends have different genotypes for warding off infection, a previously hypothesized possibility (*12*). For comparison, we conduct the same gene set analyses for strangers, and we do not find any significantly overrepresented gene sets (see SI), suggesting that the procedure does not generate false positives.

While it is possible that these identified pathways, and other pathways yet to be identified, have played an important role in recent human evolution – and, indeed, prior work shows strong positive selection "for genes related to immune response, reproduction (especially spermatogensis), and sensory perception (especially olfaction)" (*27*) – the foregoing over-representation analysis does not address whether natural selection has generally favored genotypic homophily. To test the hypothesis that homophilic SNPs are generally under recent positive selection, we use the Composite of Multiple Signals (CMS) score (*28*). This score combines signals from several measures of positive selection to create a single value that indicates the likelihood a SNP has been increasing in frequency due to selection pressure over the last 30,000 years (see SI).

In Fig.3, we show that, after correcting for correlated outcomes due to linkage disequilibrium and for varying precision in the GWAS estimates (see SI), the top 20% most homophilic SNPs have significantly higher CMS scores than the other 80% (+0.07, SE 0.02, $p = 0.003$). For comparison, note that this is about half the size of the difference in CMS scores between intragenic and intergenic SNPs (+0.15, SE 0.02), which we would expect to be large given the functional role of variants within genes relative to those between genes. In contrast, Fig.3 also shows that CMS scores are *not* significantly higher for the most homophilic SNPs in the strangers GWAS (–0.00, SE 0.02, $p = 0.86$). This suggests that the whole-genome regression model we use does not generate false positives.

Furthermore, we evaluated a model that fits the CMS score to the level of correlation in each SNP, allowing the linear relationship to be different for homophilic and heterophilic SNPs (see SI). This model (which also serves as a robustness check) shows that there is a positive and significant relationship in the friends GWAS for homophilic SNPs ($p = 0.03$). As the level of positive correlation increases, so does the expected CMS score. There is no relationship for negatively correlated (heterophilic) SNPs ($p = 0.63$). And, for comparison, there is no relationship in the strangers GWAS between genetic correlation and positive selection for either homophily ($p = 0.77$) or heterophily ($p = 0.28$). In sum, it appears that, overall, across the whole genome, the genotypes humans tend to share in common with their friends are more likely to be under recent natural selection than other genotypes.

Our analysis is explicitly genome-wide. The reason is that we expected to find a large number of weak signals of genotypic correlation. Relatedly, we examined an unavoidably limited sample of genotyped pairs of friends because such data are currently very scarce (see SI). Still, the evidence suggests that there are many SNPs that are slightly homophilic or heterophilic between friends, but we cannot yet be sure whether this means that only a few biological systems are highly correlated, or many systems are weakly correlated. Future analyses with larger samples may help to resolve this question.

It is intriguing that genetic structure in human populations may result not only from the formation of reproductive unions, but also from the formation of friendship unions. This in turn has relevance for the idea of an evocative gene-environment correlation, proposed more than 30 years ago, which suggests that a person's genes can lead one to seek out circumstances that are compatible with one's genotype (*29,30*). Our results suggest that these circumstances could include not only the physical environment but also the *social* environment, and hence the genotypic constitution of one's friends. As Tooby and Cosmides argue, "not only do individual humans have different reproductive values that can be estimated based on various cues they manifest, but they also have different association values."(*11*) People may seek out particular, convivial social environments that affect their fitness.

The existence of excess genetic similarity between friends is also relevant to the growing area of indirect genetic effects (*31*), wherein the phenotypic traits of focal individuals are influenced by the genomes of their neighbors, in a kind of "network epistasis."(*12*) In fact, our results support the idea that humans might be seen as metagenomic not just with respect to the microbes within them (*32*), but also with respect to the humans around them. It may be useful to view a person's genetic landscape as a summation of the genes within the individual and within the people surrounding the individual, just as in certain other organisms (*31,33*).

Pairs of friends are, on average, as genetically similar to one another as fourth cousins, which seems noteworthy since this estimate is net of mean ancestry and background relatedness. Acquiring friends who resemble oneself genotypically from among a group of strangers may reflect a number of processes, including the selection of particular friends or particular environments. Whatever its cause, however, the subtle process of genetic sorting in human social relationships might have an important effect on a number of other biological and social processes, from the spread of germs to the spread of information.

Insofar as the process involves the actual selection of friends, it may reflect the extended workings of some sort of kinship detector postulated in humans (*16*). One's friends, in other words, may evince a kind of functional relatedness (identity by state) – and may perhaps do so especially for particular biological systems – rather than evincing an actual relatedness (identity by descent) as in the case of kin. Forming social ties to "functional kin" who perceive or cope with the environment in a similar way to oneself can result in both individuals benefiting from each other's deliberately or accidentally created benefits ("positive externalities"); for example, if one individual builds a fire because he feels cold in the same circumstances as the other, both benefit (*11*). Genetic correlation between friends may even enhance the opportunity for natural selection to operate at the level of social groups established on a basis other than kinship; such associations have long been postulated in the theoretical evolutionary genetics literature, but there is little extant evidence (*34,35*).

Kin recognition has been shown in many vertebrates (*36*), and it is important for stabilizing cooperation and promoting inclusive fitness benefits in some species (*37*). There is suggestive evidence for some sort of kin detection system in humans as well, such that, for each individual encountered, an unspecified system may compute and update a continuous measure of kinship that corresponds to the genetic relatedness of the self to the other individual (*16*). In part, this system would be driven by objectives such as behaving altruistically towards, and avoiding sexual relations with, kin. A number of mechanisms by which kin detection might take place have been proposed, including co-residence duration monitoring; peri-natal association; and other cues, such as facial resemblance or odor. Cues of kinship may foster altruistic impulses and cooperative exchanges with individuals displaying those cues, and it is not hard to imagine that such a system might possibly be extended to preferential (active) friendship formation.

In this regard, our findings regarding homophily on certain olfactory system features are intriguing and supportive. There is evidence that olfaction plays a role in human (and other primate) kin recognition (*38,39*) and even some suggestive evidence that people are able to distinguish friends from strangers based on blind odor tests (*40,41*). The olfaction ontology in which we detect substantial homophily has some genes coding for odorant receptors; when it comes to choosing friends, perhaps a shared sense of smell is important, and the way odor messages are received, rather than just sent, may be key. Olfaction is also connected to other processes, such as emotional contagion and communication, and to the avoidance of inappropriate ingestions; these too may benefit from the synergistic presence of genotypically similar others.

The implications of the finding regarding homophily on genes related to linoleic acid metabolism are unclear. Linoleic acid is a precursor for substances involved in a broad range of important bodily processes (ranging from adipocyte function to bone formation to the regulation of gene expression) (*42*), and the component genes in the pathway are related to the metabolism of cholesterol, steroids, and various ingested substances, though it is intriguing that linoleic acid compounds might be used by moths as pheromones (*43*). Possibly, this pathway is related to the restrained consumption or the specific metabolism of various foodstuffs, traits for which homophily may be advantageous and heterophily self-injurious.

The observed heterophily on an immune system ontology has interesting implications. Prior work has provided evidence of an active process contributing to genetic heterophily between mates with respect to the avoidance of similar HLA haplotypes (*44*) (though these are not part of the present gene set). In the case of friends, there may also be advantages to complementarity rather than synergy when it comes to immune system function since surrounding oneself with others who are dissimilar to oneself in this regard may be an adaptive strategy. If one is already relatively resistant to a particular pathogen, it would be best to have friends who were resistant to different pathogens, thus mitigating the inter-personal spread of both. Genes affecting the immune system do not necessarily benefit from interpersonal ties to genotypically similar individuals.

It may be possible to use an approach similar to that outlined here, but with much larger samples of friendship pairs, and perhaps coupled with the addition of an equally large number of spousal pairs, to identify the genetic basis of kin detection. The extent to which friends and spouses resemble each other could itself be taken as a phenotype, and one could imagine doing a GWAS to isolate which regions of the genome contribute to our ability to pick suitable friends and spouses.

Finally, the human evolutionary environment is not limited to the physical environment (sunshine, altitude) or biological environment (predators, pathogens), but also includes the social environment, which may itself be an evolutionary force (*45*). Our finding that positively correlated genotypes are under positive selection suggests that the genes of other people might modify the fitness advantages of one's own genes, thus affecting the speed and outcome of evolution. In particular, communication – whether involving scent, sight, or sound – may be the key to this synergy. The human capacity to collaborate not only with kin but also with unrelated members of our species may have dramatically increased the potential gains from synergy, and this shift would not only favor interactions with generally similar partners, but would also affect the overall desire to search out such partners (*10,11*). Hence, it is possible that we evolved a predilection for homophily once we started to frequently interact socially with unrelated individuals. Such an effect would especially speed up the evolution of phenotypes that are intrinsically synergistic, and this may help shed light on the observation that evolution in humans is accelerating (*46*).

**Acknowledgements**
　　Supported by grants from the National Institute on Aging (P-01 AG031093) and the National Institute for General Medical Sciences (P-41 GM103504-03). We thank Jason Boardman, David Cesarini, Chris Dawes, Jan-Emmanuel De Neve, Feng Fu, Erez Lieberman, Martin Nowak, David Rand, Zach Steinert-Threlkeld, Juan Ugalde, and Ajit Varki for helpful comments. We thank Shervin Tabrizi for sharing the CMS data with us.

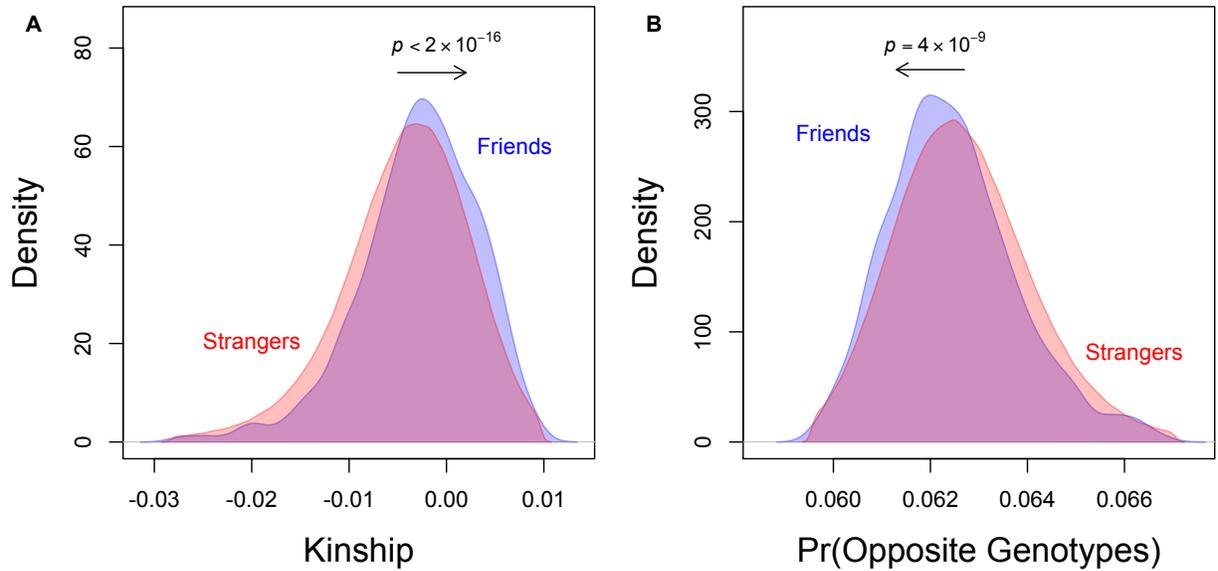

**Figure 1. Friends exhibit significantly more homophily (positive correlation) than strangers in genome-wide measures.** Overlapping density plots show that, compared to strangers, friends have (A) higher kinship coefficients and (B) lower proportions of opposite genotypes (SNPs for which neither allele is identical by state) in 1,367 friendship pairs and 1,196,429 stranger pairs observed in the same set of subjects (see SI). On average, friends have a kinship coefficient that is +0.0014 greater than friends, a value that corresponds to the relatedness of fourth cousins. *P* values are from difference-of-means tests (see SI).

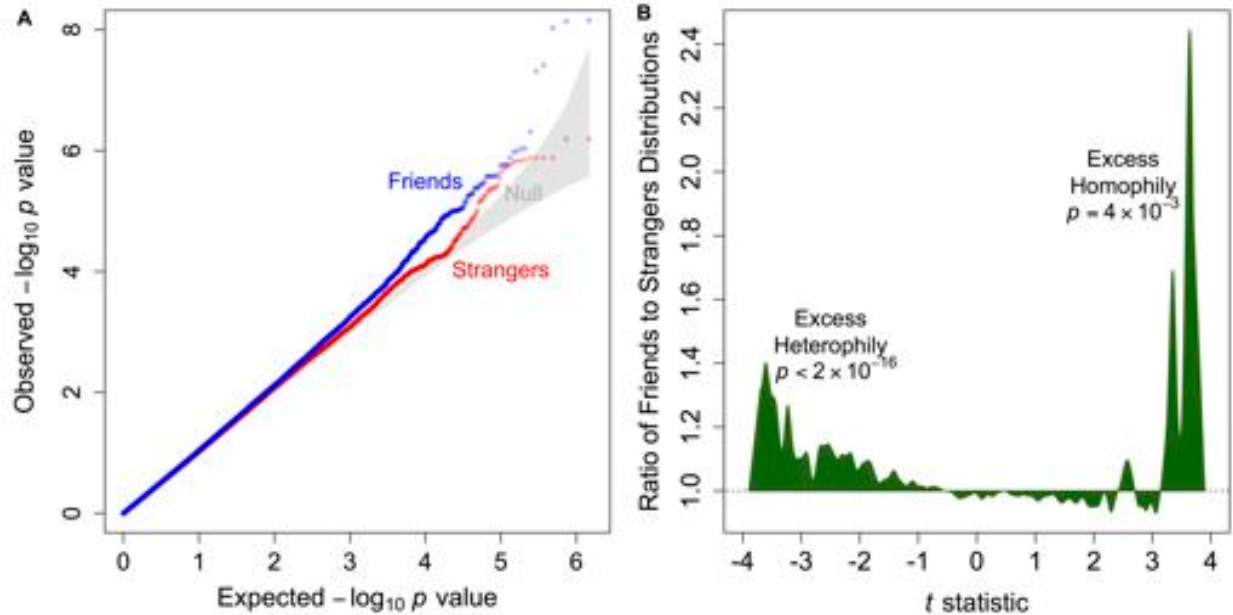

**Figure 2. Friends exhibit significantly more homophily (positive correlation) and heterophily (negative correlation) than strangers in a genome-wide association study (GWAS) with strict controls for population stratification.** (a) QQ plot of observed vs. expected *p* values from separate GWAS of genetic correlation shows more outliers for pairs of friends (blue) than pairs of strangers (red). Null distribution (gray) shows 95% confidence region for values possible due to chance. The strangers GWAS shows that some inflation is due to restricting observations to unrelated pairs of individuals, which causes genotypes to be negatively correlated on average. Over and above this baseline, the friends GWAS shows that friend pairs tend to have many markers that exhibit even lower *p* values, and this pattern is consistent with traits that are highly polygenic (*23*). (b) Distribution of *t* statistics in the friends GWAS divided by the distribution of *t* statistics in the strangers GWAS shows that friends tend to have both more heterophilic (negatively correlated) and also more homophilic (positively correlated) SNPs in the tails of the distribution. *P* values are from Kolmogorov-Smirnov tests (see SI).

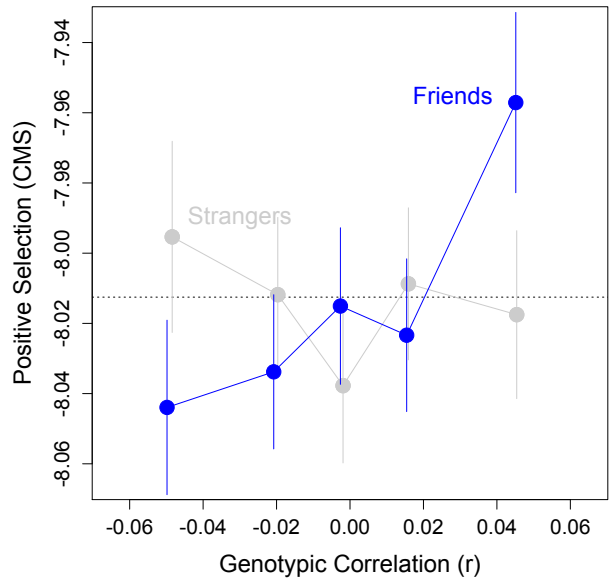

**Figure 3. Homophilic (positively correlated) SNPs are more likely to be under recent positive selection.** Plot shows mean Composite of Multiple Signals (CMS) score by SNP correlation quintile for friends (blue) and strangers (gray). Each quintile contains approximately 293,600 SNPs. Vertical lines show the standard error of the mean corrected for correlated observations due to linkage disequilibrium (see SI). For reference, the horizontal dotted line shows the mean CMS score.

|  | Friends Results | | | Strangers Results |
| --- | --- | --- | --- | --- |
|  | (1) | (2) | (3) | <none> |
| *Correlation Type* | homophily | homophily | heterophily | |
| *Set Threshold* | Top 1% | Top 25% | Top 25% | |
| *Gene Set ID* | KEGG hsa04740 | KEGG hsa00591 | GO 0002376 | |
| *Description* | Olfactory transduction | Linoleic acid metabolism | Immune system process | |
| *Number in Threshold Set* | 14 | 18 | 353 | |
| *Total Number Measured* | 355 | 29 | 1198 | |
| *Number in Gene Set* | 461 | 30 | 1562 | |
| *Percent in Threshold Set* | 3.9% | 62.1% | 29.5% | |
| *Percent Measured* | 77.0% | 96.7% | 76.7% | |
| *Z Score* | 5.05 | 4.64 | 3.55 | |
| *Fisher's Test Exact P* | $4 \times 10^{-5}$ | $2 \times 10^{-5}$ | $5 \times 10^{-4}$ | |
| *Benjamini-Hochberg Adjusted P* | 0.009 | 0.005 | 0.036 | |

**Table 1. Gene set analysis shows three genome-wide significant gene sets that are overrepresented**. Friend pairs tend to have similar gene variants (homophily) in the olfactory and linoleic acid systems, and different gene variants (heterophily) in the immune system. By comparison, the same analysis with stranger pairs shows no significant gene sets.

Supplementary Information for

# Friendship and Natural Selection

Nicholas A. Christakis[1,2] and James H. Fowler[3,4*]

[1] *Department of Medicine, Yale University, New Haven, CT 06520, USA*
[2] *Department of Sociology, Yale University, New Haven, CT 06520, USA*
[3] *Medical Genetics Division, University of California, San Diego, CA 92103, USA*
[4] *Political Science Department, University of California, San Diego, CA 92103, USA*

[*] To whom correspondence should be addressed. e-mail: fowler@ucsd.edu

## Table of Contents

## Data

The Framingham Heart Study (FHS) is a unique population-based, longitudinal, observational cohort study that was initiated in 1948 to prospectively investigate risk factors for cardiovascular disease. Since then, it has come to be composed of four separate but related cohort populations: (1) the "Original Cohort" enrolled in 1948 (*N*=5,209); (2) the "Offspring Cohort" (the children of the Original Cohort and spouses of the children) enrolled in 1971 (*N*=5,124); (3) the "Omni Cohort" enrolled in 1994 (*N*=508); and (4) the "Generation 3 Cohort" (the grandchildren of the Original Cohort) enrolled beginning in 2002 (*N*=4,095). Published reports provide details about sample composition and study design for all these cohorts; for the data we use here, follow-up occurred every 2-4 years, beginning in 1971 (*1,2,3*) As described elsewhere, we collected information identifying who was connected to whom via ties of friendship (and also ties of marriage, kinship, and so on) (*4*). This population is typical of a European-descent population in terms of SNP frequency.

In the field of social network analysis, procedures for identifying social ties between individuals are known as 'name generators' (*5*). The ascertainment of social ties in the FHS was both wide and systematic. The FHS recorded complete information about all first-order relatives (parents, spouses, siblings, children), whether alive or dead, and also about at least one 'close friend' (the set-up and question asked were "please tell us the name of a close friend, to whom you are not related… [with whom] you are close enough that they would know where you are if we can't find you").

Out of the 14,428 members of the three main cohorts, a total of 9,237 individuals have been genotyped (4,986 women and 4,251 men). Genotyping was conducted using the Affymetrix 500k array and the Affymetrix 50K supplemental array. We had data on additional genotypes for 1,345 of the participants from an Affymetrix 100K GeneChip array, so we used PLINK version 1.06 (*6*) to merge the two datasets by subject.

FHS also makes available a data set of expected genotypes. Given that this population is almost entirely white with ancestries from Europe, they have allele frequencies that are consistent with data from central European samples. Imputation of all autosomal SNPs on the publicly available phased haplotypes from the HapMap reference panel (release 22, build 36) for a Central European (CEU) population was conducted using the Affymetrix 500k and 50K arrays. Quality controls filtered out 15,586 SNPs with Hardy-Weinberg values of $p<10^{-6}$; 64,511 SNPs with missingness greater than 0.03; 45,361 SNPs with non-random missingness values of $p<10^{-9}$; 4,857 SNPs with more than 100 Mendel errors; 67,269 SNPs with frequency less than 0.01; 2 SNPs with strandedness issues when merging HapMap; and 13,394 SNPs that were not present in the HapMap.

To conduct the imputation, MACH version 1.0.15 (*7,8*) was used on a group of 200 unrelated individuals to infer model parameters. These individuals were selected by prioritizing those individuals with low missingness (worst is missingness of 0.011) and low Mendel errors (worst is 4,970 errors). Among these, 99 were females, 101 males, 164 were members of the Offspring cohort while 34 were from the Original cohort, and 2 from the Generation 3 cohort. None of these 200 individuals were identified as outliers by EIGENSTRAT (*9*). The subsequent model from MACH was then applied to all 8,481 genotyped individuals in FHS, yielding imputed dosages for 2,543,887 SNPs from the HapMap. Of these, 3,694 SNPs were "corrected" by MACH to be monomorphic (even though the minor allele frequency of these genotyped SNPs was greater than 0.01). For additional information about the procedure, see De Bakker (*10*).

There are 1,932 unique subjects who are in one or more of 1,367 friend pairs observed in the Framingham Heart Study for which genetic data is available (friend pairs observed at more than one wave of the study are only counted once). We count as a friendship any pair for which at least one person named the other person as a friend. Table S1 shows summary statistics for this sample.

## Unusual Nature of the Current Data, and Availability

To our knowledge, the dataset we develop and analyze here, which has (1) information on friendship ties for (2) a large number of people who were also (3) genotyped across the *whole genome*, is presently the only such dataset that is available anywhere. As described above, we combined newly acquired social network data (ascertaining friendship ties) with the genotyping information collected by FHS; and we have made these data available for use by others at dbGap: http://www.ncbi.nlm.nih.gov/projects/gap/cgi-bin/study.cgi?study_id=phs000153.v6.p5.

In the future, additional datasets with these features may become available and allow for further exploration of the phenomena of interest here. For example, the National Longitudinal Study of Adolescent Health (http://www.icpsr.umich.edu/icpsrweb/ICPSR/studies/27024) has thousands of friendship dyads and a subset of subjects is currently being genotyped; however, only a handful of SNPs have been made available so far, and we and others have already used this data to study genotypic correlation (*11,12*). Other possibilities for the future include commercial data (e.g., from 23andMe, though the current terms of their IRB approval do not allow analyses of the type performed here involving friendship links). The Personal Genome Project (PGP) may also be an avenue for additional data development, assuming meaningful links between participants exist and can be identified. And, it is possible that, as large numbers of people are genotyped as part of clinical databases in the future, the ascertainment of social network ties might be grafted onto such efforts.

Given that a completely independent replication data set is not presently available, we split our FHS sample in two and held part of this sample back for an out-of-sample replication test. As we describe below, the model fit to friend pairs in one sample can be used to predict a significant part of the variance in who is and is not friends in the second (replication) hold-out sample. This is one of several procedures we implement to evaluate the robustness of our findings and to reduce the likelihood of false positive results. Specifically, throughout, we also compare results for friends to results for strangers, which helps to show that any patterns in the friends could not have been generated due to random connections between the subjects in our study.

Finally, we reiterate that our interest here is in effects across the whole genome, and not in individual SNPs. Our dataset is small compared to recent consortium samples that are attempting to identify reliable associations of specific SNPs with specific traits. But this does not mean that genome-wide patterns are impossible to discern with such smaller datasets, as in other whole-genome investigations (*13,14*). We use the noisy, but still informative, measures across more than a million associations to test the hypotheses that:

1. friends generally exhibit positive correlation (homophily) in genotypes;
2. friends also exhibit heterophily on numerous genotypes;
3. certain sets of genes are significantly overrepresented among the most homophilic and most heterophilic genotypes;
4. consistent with theoretical predictions, homophilic SNPs are more likely to be under recent natural selection.

## Analysis of Relatedness

To estimate the pairwise relatedness between every single pair in the sample, we used the 466,608 *non-imputed* SNPs that survived quality checks described above. Measures of the *kinship* coefficient (the probability that two alleles sampled at random from two individuals are identical by state) and, separately, the probability of opposite genotypes, measured as the empirical proportion of SNPs for which neither allele was identical by state (i.e., where neither allele had the same nucleotide) were calculated using KING: Kinship-based INference for GWAS software version 1.4 (*15*). Higher values of the kinship coefficient indicate that two individuals share a greater number of genotypes in common (for friends, this is homophily) while higher values of the probability of opposite genotypes indicate that there are a greater number of markers where the two individuals are exactly opposite in their genotype (for friends, this is heterophily). Note that the kinship coefficient is equal to half the relatedness measure used in genome-wide complex trait analysis (GCTA) approaches (*16*).

To compare the distribution of genetic relatedness between friends and strangers, we created a "stranger pair" set that eliminated all pairs in the data that had any kind of social relation (friends, spouses, siblings, etc.) and from both the friend and stranger pair sets we also eliminated all pairs that the KING software identified as being genetically related. It is important to emphasize that individuals in the FHS did not always report a full list of all their kin, but *since we have the genetic information we can identify any non-named kin and remove them*.

Fig.1 in the main text shows the distributions of the kinship coefficient and probability of opposite genotypes for both friends and strangers. In these distributions, extreme positive relatedness may indicate true familial relationships, though these relationships would be distant since we already removed relatives. On the other hand, extreme negative relatedness may indicate that some of the pairwise comparisons are between individuals from genetically distinct populations (*15*). Since we are interested in the genetic similarity of individuals within a group from the same population who are not family members, we removed all values in both the friend pair set and the stranger pair set that were below the first percentile of the stranger pair set (kinship < –0.0283, Pr(opposite genotypes) < 0.0595) and above the 99$^{th}$ percentile of the stranger pair set (kinship > 0.0098, Pr(opposite genotypes) > 0.0671). This left 1,196,429 stranger pairs in the sample. These thresholds are arbitrary, but other thresholds yielded similar results.

Difference of means tests comparing the friends and strangers distributions suggest that friends are generally more homophilic than strangers. Kinship was higher in the friends (+0.0014, $t = 8.7$, $p < 2 \times 10^{-16}$) and the probability of opposite genotypes was lower (–0.0002, $t = 5.9$, *p* = 4 x 10$^{-9}$).  For comparison, this difference in the friends and strangers is roughly the degree of relatedness one would expect in fourth cousins, which have an expected kinship coefficient of 0.0010.

## Genome-Wide Association Study

We re-emphasize that our focus is not on individual SNPs here, but rather on the entire genome. We use the individual SNP results from a genome-wide association study (GWAS) to characterize genome-wide patterns.

Prior to conducting the GWAS, we removed 460 of the 1,367 friendship pairs with a kinship coefficient greater than zero.  We did this for two reasons: (1) we wanted to ensure that association in friend genotypes is *not*, in fact, due to people tending to make friends with distant relatives; and (2) we wanted to set aside a hold-out replication sample in which we could conduct of the capacity of the GWAS to predict out-of-sample which pairs of individuals are friends. This procedure left 907 friend pairs.  One side-effect of this conservative approach is that it biases the GWAS against finding positively correlated genotypes since the remaining friends are on average slightly negatively related (mean kinship coefficient = –0.003).

Although the individuals in the Framingham Heart Study are almost all of European ancestry, population stratification has been shown to be a concern even in samples of European Americans (*17*).  Therefore, we used EIGENSTRAT (*9*) to calculate the first ten principal components of the subject-gene matrix.  None of our subjects were classified as outliers, defined as individuals whose score is at least six standard deviations from the mean on one of the top ten principal components.  We also included the values from these components in the regression:

$$E\left[Y^{ego}\right] = \alpha + \beta_1 Y^{alter} + \sum_{i=1}^{10} \gamma_i z_i^{ego} + \sum_{i=1}^{10} \gamma_i z_i^{alter}$$

where $Y^{ego}$ is the ego's (subject's) genotype, $Y^{alter}$ is the alter's (friend's) genotype, and $z_i^{ego}$ and $z_i^{alter}$ are the *i*th principle components of the ego's and alter's genome-wide genotypes.  Note that since $Y^{ego}$ and $Y^{alter}$ have the same distribution (and therefore the same standard error), $\beta_1$ can be interpreted as a correlation coefficient *r* that has been adjusted for controls.  Note that an alternative model including sex as a control yielded identical results.

We calculated robust standard errors using multiway clustering (*18*) to account for correlated observations on subject ID to account for multiple friendships involving the same person.  Since there will be some sampling variation in a set of individuals arbitrarily assigned to be "ego," for each pair, we entered two observations – one for ego's genotype as the dependent variable and one for alter's genotype as the dependent variable – and we corrected for this double-counting by also clustering standard errors on pair ID.  So, to summarize, we clustered standard errors on *both* subject and pair ID.

We conducted this regression on each SNP using expected genotypes for each of the 1,468,013 SNPs with (1) minor allele frequency greater than 0.10, (2) nonzero variation in the number of minor alleles across observations, (3) Hardy-Weinberg values of $p < 10^{-6}$, and (4) non-missing values for the Composite of Multiple Signals (CMS) score obtained from Shervin Tabrizi and Pardis Sabeti (*19*). Analysis was conducted on the Gordon supercomputer (*20*) at the San Diego Super Computer Center.

For comparison, we repeated the GWAS for a sample of 907 stranger pairs with kinship ≤ 0, randomly drawn from the full set of 1,196,429 stranger pairs.

QQ plots of observed vs. expected negative $\log_{10} p$ values from the GWAS in Fig.2a of the main text show a small amount of inflation in test statistics. Based on the median $\chi^2$ statistic, the estimated variance inflation factor (*21*) for friends is $\lambda = 1.046$, which is modest and may be due in part to the fact that the genotypic correlation we observe is highly polygenic, which is known to cause such inflation (*22*). More than half of this inflation may be due to generalized homophily and/or heterophily as demonstrated in Fig.1 of the main text rather than population stratification or model misspecification, since the baseline variance inflation factor for strangers is only $\lambda = 1.020$. Note also that we would expect there to be at least some variance inflation since the restriction to pairs with kinship ≤ 0 means that the average pair of genotypes exhibit slight negative correlation (–0.003). This would tend to generate a higher number of lower-than-expected *p* values in negatively correlated SNPs.

To see whether the GWAS of friends generates smaller *p* values than the GWAS of strangers because of homophily (positive correlation) or heterophily (negative correlation), we compare *t* statistics from the two GWAS in Fig.2b of the main text. In that figure, we divide the estimated density of the friends distribution by the estimated density of the strangers distribution at 1000 equally spaced points across the range of observed *t* statistics to create a ratio distribution. Values greater than one suggest parts of the distribution where the likelihood of a certain *t* statistic are higher for friends than for strangers. Fig.2b shows that this ratio is high for both large positive and large negative *t* statistics, suggesting friends exhibit both excess homophily and excess heterophily compared to strangers. A Kolmogorov-Smirnov test of the difference in the distribution of negative *t* statistics between friends and strangers is significant ($D = 0.013$, $p < 2 \times 10^{-16}$). The difference in the distribution for positive *t* statistics is also significant ($D = 0.003$, $p = 4 \times 10^{-3}$).

We emphasize that our goal in this paper is not to identify specific SNPs that are correlated. Although we are likely to be underpowered to have confidence in the identification of particular SNPs (*23*), and although our focus here is on genetic correlation across the whole genome and not on particular SNPs, for completeness we show a Manhattan plot in Fig.S1 for SNPs with $p < 0.05$ for the GWAS of correlated genotypes between friends. Five SNPs are at or near the Bonferroni-corrected threshold for genome-wide significance ($p = 3.4 \times 10^{-8}$), three in the *CCNJL* gene (rs6874570, $p = 4 \times 10^{-8}$; rs4921270, $p = 7 \times 10^{-9}$; and rs6875660, $p = 7 \times 10^{-9}$), and two in the intergenic region between the *ZDHHC21* and *NFIB* genes (rs7850284, $p = 7 \times 10^{-9}$; and rs4741404, $p = 7 \times 10^{-9}$). All five of these SNPs are heterophilic (negatively correlated between friends). But again we caution that we do not expect these particular markers to survive multiple testing in larger data sets. Our goal here is to characterize genome-wide patterns and not specific markers.

For comparison, in Fig.S2 we show a Manhattan plot for the GWAS for correlated genotypes between strangers. No SNPs achieve genome-wide significance after adjusting for multiple testing (the most significant SNP is $p = 6 \times 10^{-7}$). This suggests that the excess markers with low *p* values in the friends GWAS are not merely due to chance.

## An Out-of-Sample Replication Test of the "Friendship Score"

To ascertain whether our GWAS of correlated genotypes between friends can predict outcomes out of sample, we first identified an appropriate group for testing. Prior to conducting the GWAS analysis, we removed 460 friend pairs from who had a kinship coefficient greater than 0 to ensure that we were not analysing friendships based on distant family relationship. Of these, 458 have a kinship score that is less than 0.01. We included these in our sample and we also included 458 pairs drawn at random from the stranger sample who had a kinship coefficient matched exactly to each friend pair (to control for overall relatedness). This means the total sample in this analysis has 916 pairs, half of whom are friends and half of whom are strangers.

For each SNP in the friends GWAS, we create a similarity score that is equal to 1 if the pair has two alleles identical by state, 0 if the pair has one allele identical by state, and –1 if the pair has no alleles identical by state. We then multiply the similarity score by the genotypic correlation coefficient estimated in the friends GWAS for that SNP (recall that this GWAS was originally fit on an independent sample). For each subject, we then created a "friendship score" which is the sum of these values at each SNP. We standardized these scores by subtracting the mean score and dividing by the standard error of the scores. If this score predicts friendship, then the out-of-sample friend pairs should have higher values than the out-of-sample stranger pairs.

In Table S2 we regress an indicator variable (1 = friends pair, 0 = strangers pair) on the friendship score for our 916 out-of-sample observations in the hold-out replication sample. The results show that a one standard deviation increase in the friendship score increases the probability that a pair is friends by about 6% ($p = 2 \times 10^{-4}$) and explains about 1.4% of the variance. This is similar to the variance explained using the best currently-available genetic scores for schizophrenia and bipolar disorder (0.4% to 3.2%) (*24*) and body-mass index (1.5%) (*25*), and we expect that GWAS on larger samples of friends may help to improve precision and therefore increase variance explained. A separate logit model yielded similar results (available on request).

Note that since the distribution of relatedness in the friends and strangers is, by design, identical in this out-of-sample replication test, overall relatedness cannot explain the results. Instead, they are driven by specific parts of the genome that exhibit homophily or heterophily in the original friends GWAS.

## Gene-Based Association Tests

We used VEGAS (*26*) version 0.7.30 to conduct two gene-based association tests on the likelihood that the set of SNPs within 50 kilobase pairs of each of 17,413 genes exhibited (1) homophily or (2) heterophily. Since these tests are directional, we could not simply use *p* values from the regression because they test the null hypothesis of no association. Very low *p* values might mean that a given SNP is highly homophilic if the correlation is positive or, alternatively, highly heterophilic if the correlation is negative. Therefore, we used a different approach to generate *p* values for each SNP.

For the gene-based association test of homophily, we used the *t* statistic from each SNP regression to estimate the probability that $t > 0$ using a one-tailed test. To be clear, this means that an individual SNP with a *t* statistic of 0 has a $p = 0.5$ chance of being homophilic, and $t = 2$ implies $p = 0.98$, for example. We then repeated this procedure for the gene-based association test of heterophily using a one-tailed test for each SNP to estimate the probability that $t < 0$. The distribution of *p* values in both cases is approximately uniform.

VEGAS aggregates the *p* values for all the SNPs within 50kb of a gene and incorporates information about linkage disequilibrium from the HapMap2 CEU sample to simulate the likelihood that the observed *p* values could have been generated by chance for that gene.

Although no gene in this analysis achieved genome-wide significance when correcting for multiple testing, we list in Tables S3a-c the top 1% of genes (174 genes in total) that exhibited the greatest likelihood of being homophilic and in Tables S4a-c we list the top 1% of genes (174 genes in total) that exhibited the greatest likelihood of being heterophilic. The top hit for homophily is *ZNF620* ($p = 6 \times 10^{-4}$) and the top hit for heterophily is *PQLC1* ($p = 4 \times 10^{-5}$) but neither of these achieves a conservative test of genome-wide significance since the Bonferroni-corrected significance threshold for 17,413 gene-based tests is $p = 3 \times 10^{-6}$.

## Gene Set Analysis

Although no single gene achieved gene-wide significance, the significant homophily present across the genome and evidence that friends exhibit significantly more genetic homophily and significantly more heterophily than strangers suggest that there are many genes with low levels of correlation. We therefore conducted a gene-set analysis of the most homophilic and most heterophilic genes to see if they were over-represented in any functional sets of genes.

Gene set analysis (also called pathway, gene ontology, or overrepresentation analysis) is still being developed and there are currently many software packages and approaches (*27*). Here, we use GO-ELITE version 1.2 beta (*28*) to conduct the analysis and focus on two gene set databases, KEGG pathways (*29*), which contains thousands of gene sets, and GOSlim (*30*), which is based on the top-level gene sets from the gene ontology hierarchy. KEGG has many gene sets with a small number of genes and GOSlim has fewer gene sets that tend to be composed of a larger number of genes.

GO-ELITE takes as input a "test set" of genes given by the user and tests the hypothesis that they are overrepresented in each of the gene sets in a database relative to a "reference set" of genes, which encompasses all genes included in the gene-based association test (*28*). *P* values are calculated for each gene set, and a Benjamini-Hochberg adjustment (*31*) is applied to correct for testing multiple sets using a Fisher exact test (nearly identical results are achieved when using a permutation test).

An open question is how many genes to include in the tested gene set. Researchers typically choose a few hundred (*27*), but for traits that are suspected to be polygenic, they recommend including 5% or 25% of genes from a GWAS with strongest association (*32*). We therefore used the gene-based *p* values generated by VEGAS (described above) to rank all genes and created three gene sets for testing, a top 1% set ($N = 174$ genes), a top 5% set ($N = 871$ genes), and a top 25% set ($N = 4,353$ genes).

We did this separately for the gene-based test of homophily and the gene-based test of heterophily, giving us 3 x 2 = 6 gene sets. We then tested for over-representation in two databases (KEGG (*29*) and GOSlim (*30*)) for 6 x 2 = 12 total over-representation analyses. We restricted analyses to terms associated with at least 3 genes. Table S5 shows sample sizes for test and reference gene sets for all analyses.

Table 1 in the main text shows the three gene sets that are significantly overrepresented from these analyses after correcting for multiple testing.

One potential concern is that we do not adequately control for multiple testing across gene sets, databases, and correlation types. We therefore repeated the full procedure using the gene-based *p* values calculated from the GWAS of correlated genotypes between *strangers*. Out of these 12 tests, no term in either database survived tests for multiple correction. This suggests that the procedure we have employed is unlikely to generate false positives due to multiple analyses.

Finally, we explored the position of homophilic genes within gene-gene interaction networks for each significantly overrepresented gene set using the program Cytoscape (*33,34*). Ties between genes were ascertained using the BioGRID database version release 3.2.95 (*35*). Fig.S5 shows the olfactory transduction network and highlights the genes that showed signs of homophily.

## Relationship Between Genetic Correlation and Positive Selection

To test the hypothesis that homophilic SNPs are under recent positive selection, we use the Composite of Multiple Signals (CMS) score obtained from Shervin Tabrizi and Pardis Sabeti (*36*). This score combines signals from several measures of positive selection, including the iHS test (*37*), the XP-EHH test (*38*), the ΔiHH test (*37*), the FST test (*39*), and the ΔDAF test (*36*). Scores were available for 3,179,944 SNPs, of which 1,468,013 SNPs were common (minor allele frequency > 0.1), and also present in the imputed Framingham data after the quality checks described above.

In Fig.3 of the main text and Table S6 we assign all SNPs to quintiles based on their measure of correlation from the GWAS, with the first quintile being the most negatively correlated (mean $r = -0.05$) and the fifth quintile being the most positively correlated (mean $r = 0.05$). We then use these quintile assignments to see if there is a pattern in the raw data. The results show that the mean CMS score is higher for the top quintile than it is for the other quintiles, but we need a measure of uncertainty to determine the significance of this difference.

The means and standard errors shown in Fig.3 and Table S6 are derived from a model that regresses the CMS score on a set of five fixed effects, one for each quintile. To account for different levels of certainty at each SNP, we used weighted least squares (WLS) regression (*40*), with the weight of the *i*th observation set to be inversely proportional to its variance (in our case, the squared standard error from each SNP in the GWAS). This ensures that observations measured with greater precision contribute more to the estimates. To account for serial correlation in the errors resulting from similar CMS scores and similar results from the GWAS for nearby SNPs (due to linkage disequilibrium), we use the Newey and West (*41,42*) heteroskedasticity and autocorrelation consistent (HAC) covariance matrix estimators (*43,44*).

In Tables S7-S10, we show results from a SNP-based regression using the same weighted regression technique. In Tables S7 and S8 we regress the CMS score on a constant and an indicator variable that a SNP is in the highest quintile constant. Table S7 shows that the top quintile of the most homophilic SNPs between *friends* have CMS scores that are significantly higher than the other quintiles (+0.07, p=0.003). In contrast, Table S8 shows that for *strangers* there is almost no difference in CMS scores between the top quintile and other quintiles (–0.00, *p*=0.86).

In Tables S9 and S10, we regress the CMS score on a constant and two independent variables, one for positive correlation and one for negative correlation. The positive correlation variable is equal to the GWAS estimate of correlation in SNPs when it is greater than 0 and otherwise it is zero. Similarly, the negative correlation variable is equal to the GWAS estimate of correlation in SNPs when it is less than 0 and otherwise it is zero. This allows us to estimate the relationship between the CMS score and genotypic correlation separately for homophilic and heterophilic SNPs.

Table S9 shows the results of this model for the GWAS of genotypic correlation between *friends*. The Newey-West adjusted fit statistic for the model is $F(2,1468012) = 3.23$, $p = 0.04$, suggesting there is a significant relationship between genetic correlation and the CMS score. Moreover, the Newey-West adjusted standard errors in Table S6 show that the CMS score is significantly higher for SNPs that are more positively correlated between friends (Newey-West adjusted $p = 0.03$). In contrast, there is no relationship for negatively correlated (heterophilic) SNPs ($p = 0.63$), suggesting that homophilic SNPs are driving the significance in model fit.

Table S10 shows the results of the same model for the GWAS of genotypic correlation between *strangers*. Here, the Newey-West adjusted model fit statistic is $F(2,1468012) = 0.67$, $p = 0.51$, and neither homophily ($p = 0.28$) nor heterophily ($p = 0.54$) is associated with the CMS score. These results suggest that the results for friends above are not false positives.

# Figures

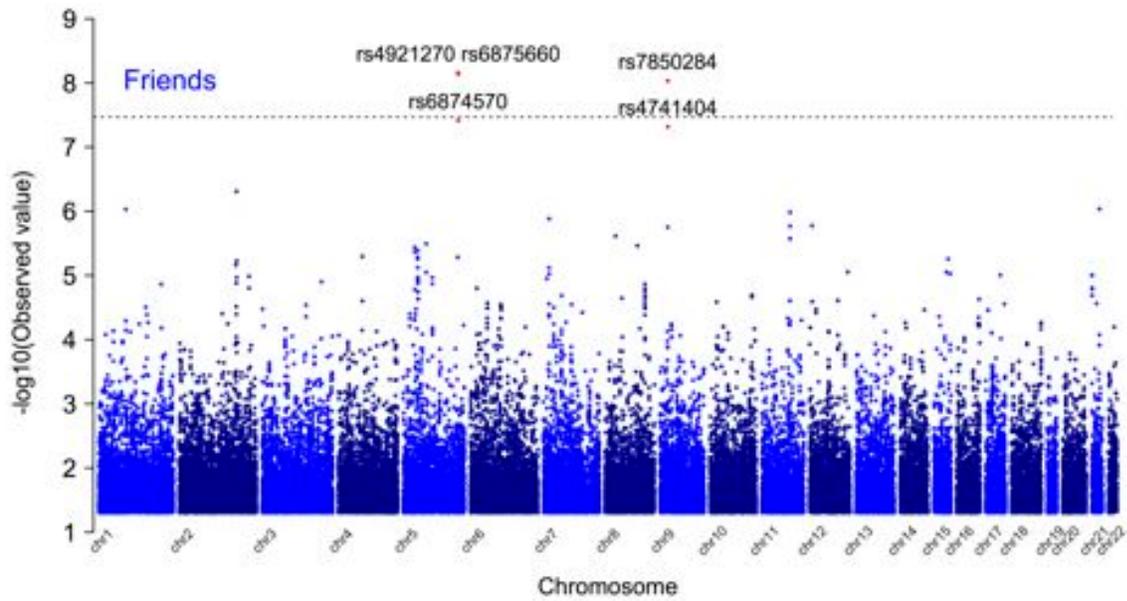

**Figure S1. Manhattan plot** showing 5 SNPs that are near or above the Bonferroni-corrected threshold for genome-wide significance (horizontal dotted line, $p = 3.4 \times 10^{-8}$) in the GWAS of correlated genotypes between friends.

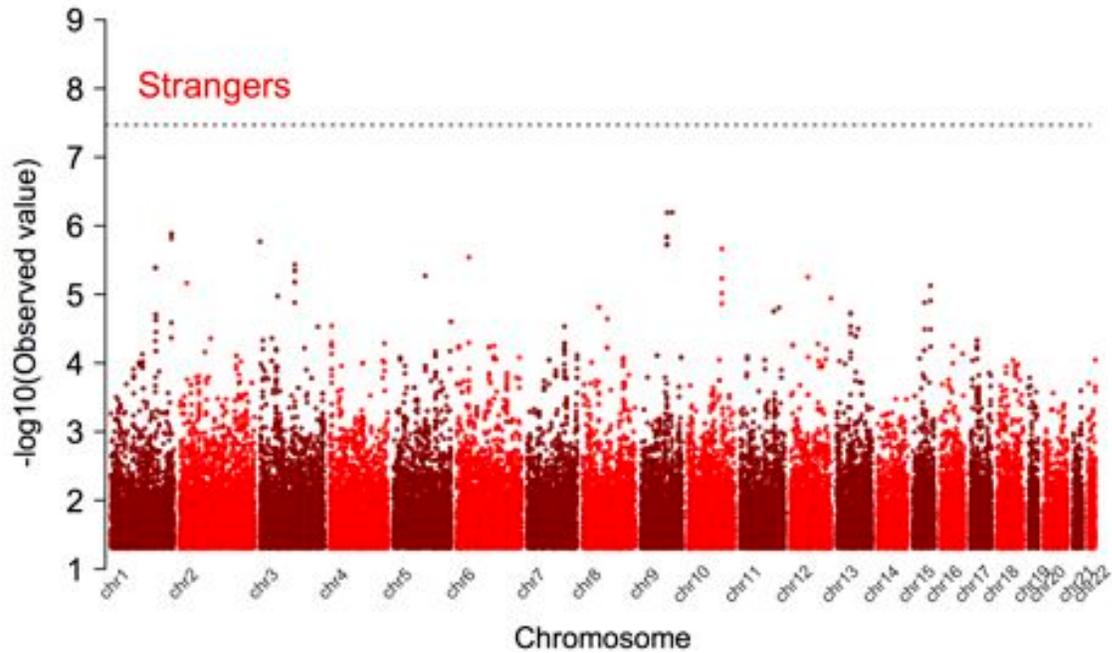

**Figure S2. Manhattan plot** showing no SNPs that are near the Bonferroni-corrected threshold for genome-wide significance (horizontal dotted line, $p = 3.4 \times 10^{-8}$) in the GWAS of correlated genotypes between strangers (minimum $p = 8 \times 10^{-7}$).

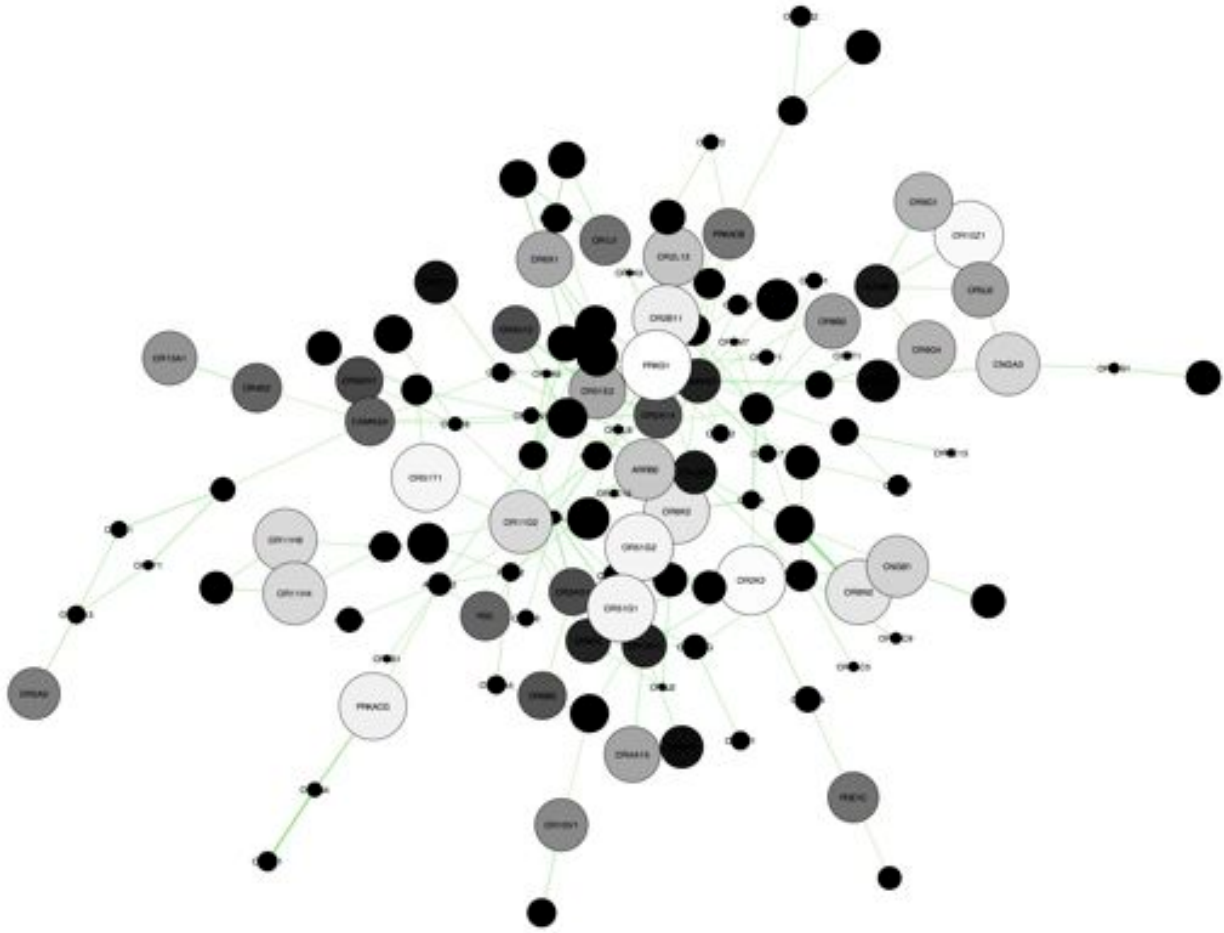

**Figure S3. Gene-gene interactions in the olfactory transduction gene set and the location of homophilic genes in this network.** Network ties were ascertained using gene interaction data available via Cytoscape 3.0. Node size is inversely proportional to gene-based *p* value, and node shade is also lighter for more significant associations. Lighter, bigger nodes indicate genes within this gene set that are more homophilic between friends.

# Tables

|  | Percent |
|---|---|
| *Female* | 56.3 |
| *Male* | 43.7 |
| *Subjects With 1 Friend* | 49.5 |
| *Subjects With 2 Friends* | 14.8 |
| *Subjects With 3 Friends* | 5.0 |
| *Subjects With 4 Friends* | 1.0 |
| *Subjects With 5 Friends* | 0.3 |
| *Subjects With 6 Friends* | 0.1 |
| *Female-Female Pairs* | 46.3 |
| *Male-Male Pairs* | 33.3 |
| *Opposite Sex Pairs* | 20.4 |

**Table S1. Summary Statistics for Friend Sample**. $N = 1,932$ unique subjects in 1,367 friend pairs.

|  | Coefficient | Standard Error | t Statistic | p Value |
|---|---|---|---|---|
| **_Friendship score_** | **0.06** | **0.02** | **3.70** | **0.0002** |
| _Constant_ | 0.50 | 0.02 | 30.46 | 0.0000 |

**Table S2. The "friendship score" based on a GWAS of correlated genotypes between friends predicts friendship in a hold-out replication sample.** Ordinary Least Squares (OLS) regression of indicator variable (1 = friend pair, 0 = stranger pair) on "friendship score" calculated as the mean over all SNPs of the product of allelic similarity in the pair and the coefficient from a GWAS of genotypic correlation between friends in an independent sample. $N$ = 916 pairs. Adjusted $R^2$ = 0.014. A logit regression yields similar results.

**Table S3a. Top Gene-Based Association Test Results from GWAS of Homophily**

| Gene | Chr | Start | Stop | Best SNP | SNP p value | # SNPs | Test Stat | Gene p value |
|---|---|---|---|---|---|---|---|---|
| ZNF620 | 3 | 40522533 | 40534042 | rs9882171 | 6 E-04 | 14 | 127 | 6 E-04 |
| ZNF619 | 3 | 40493640 | 40504881 | rs9882171 | 6 E-04 | 23 | 231 | 1 E-03 |
| COX19 | 7 | 971011 | 981761 | rs7792825 | 5 E-04 | 29 | 162 | 1 E-03 |
| RPL14 | 3 | 40473804 | 40478863 | rs9882171 | 6 E-04 | 36 | 352 | 2 E-03 |
| C5orf23 | 5 | 32824701 | 32827576 | rs1147225 | 2 E-05 | 95 | 452 | 2 E-03 |
| CYP2W1 | 7 | 989360 | 995802 | rs7792825 | 5 E-04 | 32 | 174 | 2 E-03 |
| ALDH1L1 | 3 | 127305097 | 127382175 | rs9862438 | 2 E-04 | 266 | 1118 | 2 E-03 |
| C1QL2 | 2 | 119630288 | 119632941 | rs10496557 | 2 E-05 | 82 | 315 | 2 E-03 |
| C9orf131 | 9 | 35031101 | 35035986 | rs10972278 | 2 E-04 | 33 | 213 | 2 E-03 |
| ENTPD3 | 3 | 40403676 | 40445114 | rs9882171 | 6 E-04 | 68 | 613 | 2 E-03 |
| OR10H1 | 19 | 15778816 | 15779936 | rs1821301 | 2 E-04 | 34 | 165 | 3 E-03 |
| SMPD3 | 16 | 66949730 | 67039905 | rs918788 | 2 E-03 | 132 | 611 | 3 E-03 |
| AMTN | 4 | 71418886 | 71433048 | rs17732035 | 3 E-06 | 47 | 237 | 3 E-03 |
| OR10K1 | 1 | 156701975 | 156702917 | rs1157525 | 3 E-03 | 82 | 508 | 3 E-03 |
| JAM3 | 11 | 133444029 | 133526859 | rs470500 | 5 E-04 | 93 | 536 | 3 E-03 |
| OR10H5 | 19 | 15765858 | 15766806 | rs12984671 | 2 E-04 | 47 | 236 | 3 E-03 |
| OR10R2 | 1 | 156716291 | 156717299 | rs1157525 | 3 E-03 | 87 | 552 | 3 E-03 |
| LRRC26 | 9 | 139183032 | 139184312 | rs4880094 | 4 E-03 | 1 | 8 | 4 E-03 |
| FLJ35773 | 17 | 8641205 | 8643308 | rs7207051 | 6 E-04 | 29 | 137 | 4 E-03 |
| C14orf100 | 14 | 59020913 | 59041834 | rs4898992 | 3 E-04 | 100 | 736 | 4 E-03 |
| PTPLA | 10 | 17671963 | 17699379 | rs7900258 | 1 E-03 | 84 | 342 | 4 E-03 |
| C14orf149 | 14 | 59009158 | 59020826 | rs4898992 | 3 E-04 | 98 | 687 | 4 E-03 |
| VCP | 9 | 35046064 | 35062739 | rs2252749 | 9 E-04 | 42 | 211 | 4 E-03 |
| CYP4A22 | 1 | 47375693 | 47387113 | rs12131744 | 8 E-03 | 22 | 109 | 5 E-03 |
| GPR135 | 14 | 58999992 | 59001812 | rs3736994 | 4 E-04 | 95 | 607 | 5 E-03 |
| GSK3B | 3 | 121028235 | 121295203 | rs6805251 | 7 E-04 | 126 | 516 | 5 E-03 |
| APOL1 | 22 | 34979069 | 34993523 | rs4820228 | 2 E-03 | 40 | 219 | 5 E-03 |
| SLC5A5 | 19 | 17843781 | 17866983 | rs10418352 | 6 E-03 | 11 | 48 | 5 E-03 |
| SIRPG | 20 | 1557797 | 1586425 | rs3761271 | 2 E-03 | 103 | 397 | 5 E-03 |
| NCAPD3 | 11 | 133527546 | 133599636 | rs470500 | 5 E-04 | 82 | 412 | 5 E-03 |
| IL8RB | 2 | 218698990 | 218710220 | rs7422358 | 1 E-04 | 27 | 192 | 5 E-03 |
| MUC7 | 4 | 71372524 | 71383303 | rs17732035 | 3 E-06 | 74 | 343 | 6 E-03 |
| SIGLEC11 | 19 | 55144061 | 55156241 | rs10424282 | 4 E-04 | 52 | 221 | 6 E-03 |
| PAXIP1 | 7 | 154366332 | 154425615 | rs431085 | 6 E-04 | 91 | 420 | 6 E-03 |
| RNF145 | 5 | 158516996 | 158567412 | rs2043268 | 3 E-06 | 77 | 303 | 6 E-03 |
| HSPA12B | 20 | 3661355 | 3681758 | rs4815598 | 2 E-04 | 35 | 136 | 6 E-03 |
| PIGO | 9 | 35078687 | 35086579 | rs2252749 | 9 E-04 | 47 | 193 | 6 E-03 |
| ISX | 22 | 33792129 | 33813380 | rs5755542 | 1 E-03 | 153 | 649 | 6 E-03 |
| SPINK4 | 9 | 33230195 | 33238565 | rs1702955 | 1 E-04 | 88 | 379 | 6 E-03 |
| EOMES | 3 | 27732889 | 27738789 | rs13325072 | 2 E-03 | 58 | 223 | 7 E-03 |
| CLRN1 | 3 | 152126639 | 152173476 | rs4680058 | 3 E-03 | 79 | 260 | 7 E-03 |
| FANCG | 9 | 35063834 | 35070013 | rs2252749 | 9 E-04 | 42 | 179 | 7 E-03 |
| KCNC3 | 19 | 55510576 | 55524446 | rs650829 | 5 E-03 | 22 | 85 | 7 E-03 |
| AMPH | 7 | 38389829 | 38637545 | rs4723756 | 1 E-05 | 249 | 671 | 8 E-03 |
| CST8 | 20 | 23419765 | 23424655 | rs2983299 | 4 E-04 | 45 | 235 | 8 E-03 |
| LAMC3 | 9 | 132874324 | 132958267 | rs1043169 | 3 E-03 | 106 | 326 | 8 E-03 |
| ZNF473 | 19 | 55221023 | 55243843 | rs10424282 | 4 E-04 | 54 | 282 | 8 E-03 |
| OR10K2 | 1 | 156656341 | 156657280 | rs4656284 | 3 E-03 | 60 | 324 | 8 E-03 |
| RUFY4 | 2 | 218646284 | 218663108 | rs7422358 | 1 E-04 | 26 | 124 | 8 E-03 |
| ANKS6 | 9 | 100534111 | 100598615 | rs1543506 | 8 E-04 | 67 | 264 | 8 E-03 |
| ANGPTL2 | 9 | 128889448 | 128924865 | rs1281159 | 4 E-03 | 67 | 374 | 8 E-03 |
| MYH9 | 22 | 35007271 | 35113927 | rs4820228 | 2 E-03 | 68 | 260 | 8 E-03 |
| OR10T2 | 1 | 156634935 | 156635880 | rs4656284 | 3 E-03 | 56 | 268 | 9 E-03 |
| C9orf58 | 9 | 132961732 | 132988360 | rs1043169 | 3 E-03 | 95 | 326 | 9 E-03 |
| C20orf94 | 20 | 10363950 | 10552027 | rs2179687 | 3 E-03 | 217 | 687 | 9 E-03 |
| PTCHD2 | 1 | 11461881 | 11520227 | rs10864511 | 3 E-04 | 81 | 338 | 9 E-03 |
| C20orf27 | 20 | 3682157 | 3696452 | rs4815598 | 2 E-04 | 28 | 91 | 9 E-03 |
| AKR1B10 | 7 | 133862938 | 133876700 | rs10232478 | 2 E-03 | 85 | 386 | 9 E-03 |
| VRK3 | 19 | 55171535 | 55220617 | rs10424282 | 4 E-04 | 74 | 347 | 9 E-03 |
| C20orf142 | 20 | 42368610 | 42373303 | rs6017309 | 2 E-03 | 60 | 284 | 9 E-03 |
| LCMT1 | 16 | 25030547 | 25097052 | rs277886 | 8 E-03 | 65 | 285 | 9 E-03 |
| DNAJB5 | 9 | 34979784 | 34988428 | rs10972278 | 2 E-04 | 33 | 150 | 1 E-02 |
| SAMM50 | 22 | 42682633 | 42723745 | rs2073080 | 7 E-04 | 109 | 341 | 1 E-02 |
| GYPA | 4 | 145249905 | 145281354 | rs13125760 | 6 E-05 | 18 | 81 | 1 E-02 |
| SPON2 | 4 | 1150720 | 1156980 | rs2242278 | 4 E-03 | 28 | 115 | 1 E-02 |
| SEC22A | 3 | 124403464 | 124474045 | rs3935400 | 3 E-04 | 121 | 590 | 1 E-02 |
| STOML2 | 9 | 35089888 | 35093154 | rs504082 | 9 E-04 | 47 | 163 | 1 E-02 |

**Table S3b. Top Gene-Based Association Test Results from GWAS of Homophily**

| Gene | Chr | Start | Stop | Best SNP | SNP p value | # SNPs | Test Stat | Gene p value |
|---|---|---|---|---|---|---|---|---|
| PTPLB | 3 | 124696052 | 124786614 | rs2626018 | 8 E-05 | 69 | 328 | 1 E-02 |
| AMOTL2 | 3 | 135556879 | 135576096 | rs10433423 | 7 E-04 | 55 | 187 | 1 E-02 |
| COPS3 | 17 | 17090863 | 17125316 | rs1736209 | 2 E-03 | 62 | 255 | 1 E-02 |
| CGN | 1 | 149750485 | 149777791 | rs1891593 | 6 E-04 | 51 | 160 | 1 E-02 |
| EIF1B | 3 | 40326176 | 40328919 | rs13081294 | 8 E-04 | 53 | 310 | 1 E-02 |
| GPR3 | 1 | 27591738 | 27594904 | rs12742921 | 4 E-05 | 12 | 45 | 1 E-02 |
| KIAA1539 | 9 | 35094117 | 35105893 | rs504082 | 9 E-04 | 45 | 155 | 1 E-02 |
| GDAP1L1 | 20 | 42309321 | 42342427 | rs6017309 | 2 E-03 | 75 | 307 | 1 E-02 |
| RTKN2 | 10 | 63622958 | 63698472 | rs17291653 | 6 E-04 | 127 | 524 | 1 E-02 |
| ULBP2 | 6 | 150304828 | 150312061 | rs4142628 | 6 E-03 | 21 | 74 | 1 E-02 |
| HERC6 | 4 | 89518914 | 89583272 | rs7699006 | 3 E-03 | 21 | 79 | 1 E-02 |
| TDRD3 | 13 | 59869122 | 60046012 | rs11148497 | 9 E-03 | 121 | 422 | 1 E-02 |
| TOE1 | 1 | 45577928 | 45582237 | rs9429158 | 8 E-03 | 31 | 147 | 1 E-02 |
| OR10H3 | 19 | 15713202 | 15714153 | rs12984671 | 2 E-04 | 89 | 355 | 1 E-02 |
| LOC728819 | 2 | 43755795 | 43756965 | rs10174938 | 5 E-04 | 108 | 393 | 1 E-02 |
| MTHFD2 | 2 | 74279197 | 74295932 | rs2034454 | 3 E-04 | 44 | 146 | 1 E-02 |
| OPTC | 1 | 201729893 | 201744700 | rs2096113 | 1 E-02 | 2 | 12 | 1 E-02 |
| MAP1LC3C | 1 | 240225414 | 240228998 | rs6673338 | 5 E-03 | 3 | 13 | 1 E-02 |
| MUTYH | 1 | 45567500 | 45578729 | rs9429158 | 8 E-03 | 30 | 140 | 1 E-02 |
| ANKRD34B | 5 | 79888329 | 79901854 | rs417121 | 1 E-02 | 39 | 138 | 1 E-02 |
| FLCN | 17 | 17056251 | 17081227 | rs1736209 | 2 E-03 | 56 | 243 | 1 E-02 |
| GRIN1 | 9 | 139153429 | 139183029 | rs4880094 | 4 E-03 | 10 | 42 | 1 E-02 |
| TUFT1 | 1 | 149779404 | 149822683 | rs1891593 | 6 E-04 | 49 | 150 | 1 E-02 |
| C4orf6 | 4 | 5577783 | 5580428 | rs4689259 | 3 E-03 | 53 | 232 | 1 E-02 |
| PRR6 | 17 | 16186572 | 16197537 | rs2074995 | 6 E-03 | 20 | 103 | 1 E-02 |
| NPR3 | 5 | 32747421 | 32823011 | rs1147225 | 2 E-05 | 133 | 382 | 1 E-02 |
| OR6K2 | 1 | 156936091 | 156937066 | rs373536 | 5 E-03 | 96 | 366 | 1 E-02 |
| ZNF621 | 3 | 40541379 | 40556047 | rs9882171 | 6 E-04 | 14 | 77 | 1 E-02 |
| EMX1 | 2 | 72998111 | 73015528 | rs10865394 | 8 E-03 | 29 | 98 | 1 E-02 |
| SPTA1 | 1 | 156847119 | 156923130 | rs12088990 | 2 E-03 | 129 | 427 | 1 E-02 |
| C7orf23 | 7 | 86663413 | 86686967 | rs10273826 | 3 E-03 | 45 | 205 | 1 E-02 |
| RABGAP1L | 1 | 172395256 | 173193950 | rs12072050 | 3 E-03 | 289 | 1414 | 1 E-02 |
| MGA | 15 | 39739901 | 39849433 | rs2695169 | 1 E-03 | 70 | 256 | 1 E-02 |
| EVC2 | 4 | 5615052 | 5761195 | rs9995842 | 3 E-03 | 196 | 475 | 1 E-02 |
| CYP4Z1 | 1 | 47305746 | 47356579 | rs12041262 | 1 E-03 | 33 | 172 | 1 E-02 |
| HPDL | 1 | 45565131 | 45566933 | rs9429158 | 8 E-03 | 25 | 107 | 1 E-02 |
| TPT1 | 13 | 44809303 | 44813297 | rs7987531 | 5 E-03 | 15 | 66 | 1 E-02 |
| LEFTY2 | 1 | 224190925 | 224195543 | rs2749698 | 2 E-03 | 26 | 77 | 1 E-02 |
| WSCD1 | 17 | 5914657 | 5968471 | rs6502926 | 7 E-03 | 83 | 230 | 1 E-02 |
| PRDM2 | 1 | 13903936 | 14024162 | rs1999943 | 2 E-03 | 122 | 397 | 1 E-02 |
| PNPLA3 | 22 | 42650951 | 42674781 | rs2073079 | 8 E-04 | 111 | 311 | 1 E-02 |
| CD164L2 | 1 | 27578182 | 27582392 | rs12742921 | 4 E-05 | 13 | 45 | 1 E-02 |
| OR10Z1 | 1 | 156842852 | 156843794 | rs12088990 | 2 E-03 | 68 | 256 | 1 E-02 |
| CENTA1 | 7 | 904062 | 960815 | rs7792825 | 5 E-04 | 42 | 124 | 1 E-02 |
| OR10X1 | 1 | 156815332 | 156816313 | rs12088990 | 2 E-03 | 88 | 382 | 2 E-02 |
| RAET1G | 6 | 150279706 | 150285907 | rs4142628 | 6 E-03 | 55 | 243 | 2 E-02 |
| SERPINB2 | 18 | 59705921 | 59722100 | rs1243059 | 1 E-03 | 144 | 712 | 2 E-02 |
| PLXNA1 | 3 | 128190191 | 128238920 | rs4679307 | 3 E-03 | 40 | 157 | 2 E-02 |
| HELB | 12 | 64982622 | 65018225 | rs1168309 | 2 E-02 | 85 | 352 | 2 E-02 |
| RALGPS1 | 9 | 128716873 | 129025264 | rs1281159 | 4 E-03 | 195 | 775 | 2 E-02 |
| ILKAP | 2 | 238743781 | 238777063 | rs12731 | 1 E-02 | 46 | 128 | 2 E-02 |
| USP33 | 1 | 77934261 | 77998125 | rs203229 | 5 E-03 | 33 | 171 | 2 E-02 |
| GPR52 | 1 | 172683834 | 172685306 | rs10912779 | 7 E-03 | 38 | 195 | 2 E-02 |
| HSPA14 | 10 | 14920266 | 14953746 | rs7896464 | 5 E-03 | 30 | 95 | 2 E-02 |
| OR10H2 | 19 | 15699833 | 15700862 | rs12984671 | 2 E-04 | 100 | 362 | 2 E-02 |
| XRN2 | 20 | 21231941 | 21318463 | rs6047355 | 2 E-03 | 56 | 279 | 2 E-02 |
| WDTC1 | 1 | 27433745 | 27507322 | rs12742921 | 4 E-05 | 22 | 72 | 2 E-02 |
| IL8RA | 2 | 218735812 | 218739961 | rs4674261 | 3 E-04 | 30 | 134 | 2 E-02 |
| SAC | 1 | 166045505 | 166149964 | rs2097572 | 2 E-05 | 101 | 258 | 2 E-02 |
| GTF2F2 | 13 | 44592630 | 44756239 | rs5029138 | 4 E-03 | 75 | 342 | 2 E-02 |
| PPP2R3A | 3 | 137167256 | 137349423 | rs16843509 | 1 E-03 | 93 | 367 | 2 E-02 |
| ZNF207 | 17 | 27701269 | 27721581 | rs6505294 | 2 E-03 | 24 | 120 | 2 E-02 |
| PLA2G2D | 1 | 20311020 | 20318595 | rs2020886 | 2 E-03 | 56 | 138 | 2 E-02 |
| OR6Y1 | 1 | 156783541 | 156784519 | rs10908677 | 5 E-03 | 86 | 402 | 2 E-02 |
| USP43 | 17 | 9489674 | 9573728 | rs4791863 | 2 E-03 | 69 | 210 | 2 E-02 |
| HLCS | 21 | 37045058 | 37284373 | rs1893654 | 4 E-05 | 206 | 609 | 2 E-02 |
| KLHDC6 | 3 | 129124591 | 129189204 | rs11915597 | 7 E-04 | 93 | 296 | 2 E-02 |

**Table S3c. Top Gene-Based Association Test Results from GWAS of Homophily**

| Gene | Chr | Start | Stop | Best SNP | SNP p value | # SNPs | Test Stat | Gene p value |
|---|---|---|---|---|---|---|---|---|
| *ADAMTS5* | 21 | 27212111 | 27260703 | rs438349 | 1 E-03 | 74 | 228 | 2 E-02 |
| *OR6K3* | 1 | 156953581 | 156954577 | rs373536 | 5 E-03 | 116 | 381 | 2 E-02 |
| *SLC4A3* | 2 | 220200535 | 220214946 | rs12694472 | 1 E-03 | 10 | 34 | 2 E-02 |
| *ADCY5* | 3 | 124486088 | 124650082 | rs3935400 | 3 E-04 | 134 | 447 | 2 E-02 |
| *CYP3A4* | 7 | 99192539 | 99219744 | rs651430 | 2 E-02 | 1 | 6 | 2 E-02 |
| *LCN12* | 9 | 138966588 | 138969770 | rs13301872 | 1 E-02 | 24 | 105 | 2 E-02 |
| *PLXNB2* | 22 | 49055534 | 49075336 | rs1555048 | 1 E-03 | 14 | 50 | 2 E-02 |
| *SLC25A44* | 1 | 154430522 | 154449211 | rs2241109 | 6 E-03 | 35 | 111 | 2 E-02 |
| *EVX1* | 7 | 27248688 | 27252717 | rs1476658 | 2 E-02 | 39 | 118 | 2 E-02 |
| *SNRPD1* | 18 | 17446257 | 17464206 | rs2847129 | 1 E-03 | 26 | 126 | 2 E-02 |
| *SLC46A1* | 17 | 23745787 | 23757355 | rs708100 | 1 E-02 | 16 | 72 | 2 E-02 |
| *OR6N2* | 1 | 157013095 | 157014049 | rs373536 | 5 E-03 | 110 | 307 | 2 E-02 |
| *SLC6A2* | 16 | 54248056 | 54295201 | rs2134253 | 1 E-03 | 97 | 259 | 2 E-02 |
| *SERPINB10* | 18 | 59733724 | 59753456 | rs9951512 | 4 E-03 | 145 | 643 | 2 E-02 |
| *TIPRL* | 1 | 166414794 | 166437975 | rs1406822 | 4 E-03 | 50 | 166 | 2 E-02 |
| *RBP2* | 3 | 140654415 | 140678042 | rs211584 | 2 E-03 | 112 | 436 | 2 E-02 |
| *CCDC4* | 4 | 41807713 | 41849652 | rs13756 | 6 E-03 | 55 | 248 | 2 E-02 |
| *SMG7* | 1 | 181708256 | 181789949 | rs2702199 | 1 E-02 | 68 | 222 | 2 E-02 |
| *CST11* | 20 | 23379040 | 23381482 | rs2983299 | 4 E-04 | 43 | 186 | 2 E-02 |
| *WASF2* | 1 | 27604712 | 27689256 | rs12742921 | 4 E-05 | 8 | 30 | 2 E-02 |
| *RIPK2* | 8 | 90839109 | 90872433 | rs10094579 | 5 E-03 | 83 | 360 | 2 E-02 |
| *GRAMD2* | 15 | 70239201 | 70277180 | rs2034879 | 3 E-03 | 21 | 107 | 2 E-02 |
| *DSCR6* | 21 | 37300732 | 37313828 | rs1893654 | 4 E-05 | 53 | 148 | 2 E-02 |
| *AKIRIN1* | 1 | 39229503 | 39243566 | rs12032116 | 7 E-04 | 39 | 130 | 2 E-02 |
| *FBXW7* | 4 | 153461859 | 153675622 | rs1484879 | 8 E-03 | 36 | 184 | 2 E-02 |
| *ANKRD12* | 18 | 9126757 | 9275206 | rs6506649 | 4 E-03 | 85 | 292 | 2 E-02 |
| *C1orf127* | 1 | 10929119 | 10946845 | rs11121667 | 2 E-03 | 44 | 112 | 2 E-02 |
| *ATF5* | 19 | 55123785 | 55129004 | rs7246244 | 1 E-03 | 41 | 121 | 2 E-02 |
| *B4GALT1* | 9 | 33100638 | 33157356 | rs1702955 | 1 E-04 | 112 | 313 | 2 E-02 |
| *R3HDML* | 20 | 42399039 | 42413289 | rs6031507 | 4 E-03 | 52 | 152 | 2 E-02 |
| *GOLGA7* | 8 | 41467237 | 41487656 | rs10094451 | 5 E-03 | 95 | 317 | 2 E-02 |
| *SARM1* | 17 | 23723113 | 23752192 | rs708100 | 1 E-02 | 19 | 85 | 2 E-02 |
| *HNRNPF* | 10 | 43201070 | 43224702 | rs2247979 | 3 E-03 | 32 | 120 | 2 E-02 |
| *KCTD4* | 13 | 44664987 | 44673175 | rs5029138 | 4 E-03 | 33 | 177 | 2 E-02 |
| *TUBGCP3* | 13 | 112187328 | 112290482 | rs9550148 | 1 E-02 | 120 | 434 | 2 E-02 |
| *MPZL1* | 1 | 165957831 | 166026684 | rs2097572 | 2 E-05 | 116 | 305 | 2 E-02 |
| *DMTF1* | 7 | 86619859 | 86663584 | rs10273826 | 3 E-03 | 68 | 274 | 2 E-02 |
| *EDN3* | 20 | 57308893 | 57334442 | rs1407538 | 8 E-03 | 51 | 168 | 2 E-02 |
| *TNFAIP1* | 17 | 23686912 | 23698160 | rs708100 | 1 E-02 | 14 | 66 | 2 E-02 |
| *C17orf75* | 17 | 27682501 | 27693302 | rs6505294 | 2 E-03 | 30 | 115 | 2 E-02 |

**Table S4a. Top Gene-Based Association Test Results from GWAS of Heterophily**

| Gene | Chr | Start | Stop | Best SNP | SNP p value | # SNPs | Test Stat | Gene p value |
|---|---|---|---|---|---|---|---|---|
| PQLC1 | 18 | 75763474 | 75812605 | rs1134231 | 3 E-05 | 55 | 458 | 4 E-05 |
| TXNL4A | 18 | 75833854 | 75849520 | rs4799119 | 3 E-05 | 65 | 490 | 1 E-04 |
| HMGA2 | 12 | 64504506 | 64646338 | rs2260663 | 4 E-05 | 69 | 424 | 2 E-04 |
| SSH2 | 17 | 24977090 | 25281144 | rs3115086 | 4 E-05 | 85 | 929 | 3 E-04 |
| TREML2 | 6 | 41265529 | 41276903 | rs1825411 | 1 E-05 | 85 | 552 | 4 E-04 |
| KCNG2 | 18 | 75724655 | 75760804 | rs4799099 | 7 E-05 | 40 | 239 | 4 E-04 |
| EFCAB5 | 17 | 25292811 | 25459596 | rs4294864 | 1 E-04 | 66 | 788 | 4 E-04 |
| C19orf26 | 19 | 1180946 | 1188990 | rs741765 | 5 E-04 | 1 | 12 | 4 E-04 |
| MIDN | 19 | 1199551 | 1210142 | rs741765 | 5 E-04 | 1 | 12 | 4 E-04 |
| C19orf23 | 19 | 1218469 | 1221259 | rs741765 | 5 E-04 | 1 | 12 | 5 E-04 |
| CIRBP | 19 | 1220266 | 1224171 | rs741765 | 5 E-04 | 1 | 12 | 5 E-04 |
| ATP5D | 19 | 1192748 | 1195824 | rs741765 | 5 E-04 | 1 | 12 | 5 E-04 |
| CCDC55 | 17 | 25467959 | 25537612 | rs4294864 | 1 E-04 | 59 | 593 | 5 E-04 |
| MAPK13 | 6 | 36206239 | 36215820 | rs12200998 | 9 E-05 | 29 | 162 | 5 E-04 |
| GINS2 | 16 | 84268780 | 84280081 | rs419504 | 1 E-05 | 65 | 404 | 9 E-04 |
| TREML4 | 6 | 41304039 | 41314098 | rs1825411 | 1 E-05 | 112 | 572 | 9 E-04 |
| PDIK1L | 1 | 26310912 | 26324626 | rs12088601 | 2 E-04 | 49 | 353 | 9 E-04 |
| RPL30 | 8 | 99123117 | 99126949 | rs2444891 | 3 E-04 | 64 | 332 | 1 E-03 |
| C1orf146 | 1 | 92456160 | 92483955 | rs1226176 | 1 E-04 | 27 | 188 | 1 E-03 |
| PRSS16 | 6 | 27323486 | 27332229 | rs6913724 | 7 E-05 | 40 | 237 | 1 E-03 |
| KBTBD8 | 3 | 67132092 | 67144322 | rs1010505 | 3 E-05 | 65 | 383 | 1 E-03 |
| NETO2 | 16 | 45672942 | 45735409 | rs3095622 | 1 E-03 | 6 | 36 | 1 E-03 |
| CPZ | 4 | 8645334 | 8672388 | rs3756176 | 1 E-04 | 52 | 171 | 1 E-03 |
| CGA | 6 | 87851940 | 87861543 | rs6631 | 1 E-03 | 76 | 443 | 1 E-03 |
| BTBD8 | 1 | 92318480 | 92385985 | rs547699 | 5 E-07 | 61 | 508 | 1 E-03 |
| ITFG1 | 16 | 45746798 | 46052516 | rs3095622 | 1 E-03 | 14 | 103 | 1 E-03 |
| WDR63 | 1 | 85300594 | 85371407 | rs709783 | 5 E-05 | 122 | 456 | 1 E-03 |
| RUFY1 | 5 | 178910176 | 178969625 | rs11249638 | 3 E-04 | 43 | 231 | 1 E-03 |
| STK11 | 19 | 1156797 | 1179434 | rs741765 | 5 E-04 | 2 | 13 | 2 E-03 |
| ATG3 | 3 | 113734048 | 113763175 | rs17235583 | 6 E-04 | 7 | 46 | 2 E-03 |
| PHKB | 16 | 46052710 | 46292935 | rs4966454 | 3 E-03 | 10 | 65 | 2 E-03 |
| NSMCE2 | 8 | 126173276 | 126448544 | rs4279612 | 7 E-06 | 112 | 499 | 2 E-03 |
| BCL7B | 7 | 72588621 | 72609960 | rs11983997 | 7 E-04 | 11 | 92 | 2 E-03 |
| C16orf74 | 16 | 84298624 | 84342190 | rs419504 | 1 E-05 | 92 | 440 | 2 E-03 |
| MAPK14 | 6 | 36103550 | 36186513 | rs12200998 | 9 E-05 | 63 | 311 | 2 E-03 |
| ABHD7 | 1 | 92268145 | 92301681 | rs547699 | 5 E-07 | 41 | 216 | 2 E-03 |
| CCDC34 | 11 | 27316636 | 27341371 | rs1026715 | 2 E-04 | 85 | 402 | 2 E-03 |
| GPR78 | 4 | 8633190 | 8640420 | rs3756176 | 1 E-04 | 53 | 173 | 2 E-03 |
| TRIM63 | 1 | 26250384 | 26266708 | rs17257100 | 2 E-04 | 52 | 315 | 2 E-03 |
| SYDE2 | 1 | 85395943 | 85439316 | rs709783 | 5 E-05 | 86 | 336 | 2 E-03 |
| CCDC122 | 13 | 43308488 | 43351826 | rs1813293 | 3 E-04 | 94 | 510 | 2 E-03 |
| KLF10 | 8 | 103730187 | 103737128 | rs7827122 | 3 E-04 | 73 | 278 | 2 E-03 |
| KCNJ9 | 1 | 158317983 | 158325836 | rs2753268 | 5 E-04 | 49 | 217 | 2 E-03 |
| HTR1A | 5 | 63292033 | 63293302 | rs376955 | 2 E-06 | 36 | 228 | 3 E-03 |
| TSG101 | 11 | 18458433 | 18505065 | rs4756950 | 1 E-03 | 67 | 371 | 3 E-03 |
| DYNLL2 | 17 | 53515797 | 53521810 | rs8069790 | 5 E-06 | 24 | 112 | 3 E-03 |
| CORO6 | 17 | 24965899 | 24972620 | rs3098950 | 1 E-04 | 20 | 111 | 3 E-03 |
| C18orf22 | 18 | 75895345 | 75907377 | rs4799119 | 3 E-05 | 90 | 444 | 3 E-03 |
| GRRP1 | 1 | 26358097 | 26361706 | rs12088601 | 2 E-04 | 60 | 272 | 3 E-03 |
| IGSF8 | 1 | 158327753 | 158335032 | rs2753268 | 5 E-04 | 51 | 215 | 3 E-03 |
| SERPINB6 | 6 | 2893391 | 2917089 | rs318458 | 1 E-04 | 91 | 320 | 3 E-03 |
| CCNJL | 5 | 159611248 | 159672151 | rs4921270 | 3 E-09 | 64 | 205 | 3 E-03 |
| RBM47 | 4 | 40120028 | 40326640 | rs17587454 | 8 E-05 | 95 | 262 | 3 E-03 |
| SLC6A4 | 17 | 25549031 | 25586841 | rs1042173 | 2 E-04 | 50 | 283 | 3 E-03 |
| FMO2 | 1 | 169421011 | 169448446 | rs17350523 | 9 E-04 | 96 | 430 | 3 E-03 |
| ATP1A2 | 1 | 158352171 | 158379998 | rs2753268 | 5 E-04 | 81 | 242 | 3 E-03 |
| KCNJ10 | 1 | 158274656 | 158306585 | rs2753268 | 5 E-04 | 88 | 324 | 3 E-03 |
| SH2D6 | 2 | 85515428 | 85517663 | rs7594872 | 9 E-04 | 48 | 198 | 3 E-03 |
| LGR4 | 11 | 27344083 | 27450910 | rs1026715 | 2 E-04 | 86 | 336 | 3 E-03 |
| C8orf47 | 8 | 99145925 | 99175014 | rs13278732 | 7 E-04 | 74 | 215 | 4 E-03 |
| BAZ1B | 7 | 72492675 | 72574544 | rs1178977 | 5 E-04 | 16 | 117 | 4 E-03 |
| MAB21L1 | 13 | 34946341 | 34948788 | rs2027560 | 1 E-04 | 58 | 262 | 4 E-03 |
| TMEM17 | 2 | 62581263 | 62586980 | rs17025634 | 2 E-03 | 76 | 264 | 4 E-03 |
| GRK5 | 10 | 120957186 | 121205121 | rs1556709 | 4 E-04 | 187 | 571 | 4 E-03 |
| CRISPLD2 | 16 | 83411112 | 83500615 | rs11862286 | 2 E-04 | 159 | 451 | 4 E-03 |
| EXPH5 | 11 | 107884885 | 107969570 | rs2846407 | 2 E-03 | 85 | 292 | 4 E-03 |
| FBXW11 | 5 | 171221160 | 171366482 | rs10079542 | 7 E-04 | 39 | 131 | 4 E-03 |

**Table S4b. Top Gene-Based Association Test Results from GWAS of Heterophily**

| Gene | Chr | Start | Stop | Best SNP | SNP p value | # SNPs | Test Stat | Gene p value |
|---|---|---|---|---|---|---|---|---|
| FABP6 | 5 | 159558625 | 159598307 | rs4921270 | 3 E-09 | 47 | 172 | 4 E-03 |
| AFM | 4 | 74566325 | 74588582 | rs10938078 | 1 E-03 | 63 | 376 | 4 E-03 |
| C13orf31 | 13 | 43351419 | 43366068 | rs3088362 | 3 E-04 | 86 | 428 | 4 E-03 |
| DLX2 | 2 | 172672411 | 172675724 | rs10930506 | 3 E-03 | 37 | 130 | 4 E-03 |
| KCNMB2 | 3 | 179736917 | 180044911 | rs2863186 | 4 E-04 | 354 | 1144 | 4 E-03 |
| PMM2 | 16 | 8799170 | 8850695 | rs1865806 | 1 E-04 | 84 | 319 | 4 E-03 |
| TMEM206 | 1 | 210604438 | 210654890 | rs4951588 | 7 E-06 | 83 | 346 | 4 E-03 |
| UEVLD | 11 | 18509819 | 18566857 | rs4756950 | 1 E-03 | 85 | 394 | 4 E-03 |
| C15orf37 | 15 | 78002167 | 78004249 | rs1814345 | 4 E-05 | 48 | 161 | 4 E-03 |
| ZNF593 | 1 | 26368974 | 26369951 | rs17163588 | 2 E-03 | 60 | 238 | 4 E-03 |
| CARD14 | 17 | 75766875 | 75797446 | rs12451566 | 2 E-04 | 71 | 239 | 4 E-03 |
| PRKG2 | 4 | 82228860 | 82345239 | rs6821258 | 2 E-04 | 95 | 432 | 4 E-03 |
| PABPC4L | 4 | 135336938 | 135342353 | rs2421229 | 1 E-03 | 63 | 375 | 4 E-03 |
| KIAA0182 | 16 | 84204424 | 84267313 | rs1053328 | 3 E-03 | 55 | 270 | 5 E-03 |
| FBXO2 | 1 | 11631034 | 11637326 | rs7550768 | 1 E-03 | 23 | 113 | 5 E-03 |
| TBL2 | 7 | 72621209 | 72630949 | rs11983997 | 7 E-04 | 14 | 96 | 5 E-03 |
| FBXO44 | 1 | 11637018 | 11645971 | rs7550768 | 1 E-03 | 22 | 104 | 5 E-03 |
| KCNAB1 | 3 | 157321030 | 157739621 | rs17170 | 6 E-04 | 355 | 1043 | 5 E-03 |
| CAPG | 2 | 85475381 | 85491187 | rs7594872 | 9 E-04 | 52 | 190 | 5 E-03 |
| TAS2R1 | 5 | 9682108 | 9683463 | rs7712796 | 3 E-04 | 71 | 228 | 5 E-03 |
| MUC17 | 7 | 100450083 | 100488860 | rs10231438 | 3 E-04 | 45 | 195 | 5 E-03 |
| FBXO6 | 1 | 11646767 | 11656996 | rs7550768 | 1 E-03 | 19 | 92 | 5 E-03 |
| SLC30A2 | 1 | 26237100 | 26245191 | rs12067677 | 1 E-03 | 61 | 295 | 5 E-03 |
| CCNH | 5 | 86725837 | 86744592 | rs10065414 | 3 E-04 | 20 | 123 | 5 E-03 |
| PARD6G | 18 | 76016105 | 76106388 | rs9956207 | 2 E-03 | 79 | 294 | 5 E-03 |
| WASF1 | 6 | 110527714 | 110607900 | rs4339500 | 9 E-04 | 29 | 195 | 5 E-03 |
| TMEM186 | 16 | 8796821 | 8798991 | rs1865806 | 1 E-04 | 59 | 237 | 5 E-03 |
| CLEC14A | 14 | 37793069 | 37795325 | rs1958468 | 2 E-04 | 58 | 199 | 5 E-03 |
| C7orf45 | 7 | 129634953 | 129644733 | rs10282425 | 8 E-03 | 39 | 163 | 5 E-03 |
| KIAA0406 | 20 | 36044836 | 36095247 | rs2273349 | 1 E-04 | 68 | 360 | 5 E-03 |
| C10orf141 | 10 | 128823679 | 128884412 | rs11016125 | 2 E-04 | 128 | 622 | 5 E-03 |
| CCKAR | 4 | 26092115 | 26101140 | rs2968719 | 1 E-04 | 59 | 203 | 5 E-03 |
| ABAT | 16 | 8675944 | 8785933 | rs1865806 | 1 E-04 | 140 | 394 | 5 E-03 |
| ST8SIA2 | 15 | 90738143 | 90812962 | rs4777980 | 2 E-03 | 135 | 381 | 5 E-03 |
| MOBKL3 | 2 | 198088564 | 198125760 | rs2340687 | 2 E-03 | 32 | 170 | 5 E-03 |
| SULT1A2 | 16 | 28510766 | 28515892 | rs4788073 | 4 E-03 | 16 | 88 | 5 E-03 |
| PPP2R5A | 1 | 210525501 | 210601828 | rs4951588 | 7 E-06 | 112 | 466 | 6 E-03 |
| SULT1B1 | 4 | 70627274 | 70661019 | rs12507510 | 2 E-03 | 76 | 356 | 6 E-03 |
| THNSL2 | 2 | 88250949 | 88267261 | rs2970924 | 2 E-03 | 93 | 357 | 6 E-03 |
| SGSH | 17 | 75797673 | 75808794 | rs12451566 | 2 E-04 | 56 | 194 | 6 E-03 |
| MYO10 | 5 | 16715015 | 16989385 | rs25910 | 2 E-05 | 239 | 578 | 6 E-03 |
| VWC2 | 7 | 49783802 | 49922684 | rs692270 | 2 E-04 | 147 | 593 | 6 E-03 |
| GNA13 | 17 | 60437294 | 60483216 | rs8082708 | 2 E-03 | 13 | 71 | 6 E-03 |
| FBXO38 | 5 | 147743738 | 147802592 | rs17720155 | 2 E-03 | 96 | 581 | 6 E-03 |
| TMEM209 | 7 | 129591790 | 129632574 | rs10282425 | 8 E-03 | 43 | 162 | 6 E-03 |
| FZD9 | 7 | 72486044 | 72488386 | rs1178977 | 5 E-04 | 10 | 58 | 6 E-03 |
| SULT1A1 | 16 | 28524416 | 28542367 | rs4788073 | 4 E-03 | 14 | 73 | 6 E-03 |
| GYPE | 4 | 145011468 | 145046166 | rs6834494 | 3 E-03 | 16 | 90 | 6 E-03 |
| LAPTM4A | 2 | 20095893 | 20114926 | rs10016316 | 6 E-04 | 74 | 302 | 6 E-03 |
| MRPS24 | 7 | 43872681 | 43875670 | rs1181531 | 4 E-04 | 39 | 157 | 6 E-03 |
| ROS1 | 6 | 117716222 | 117853711 | rs13192511 | 1 E-03 | 164 | 470 | 6 E-03 |
| C20orf77 | 20 | 36095361 | 36154180 | rs2273349 | 1 E-04 | 88 | 440 | 7 E-03 |
| IRF2 | 4 | 185545869 | 185632720 | rs10000856 | 6 E-05 | 100 | 249 | 7 E-03 |
| LRIG2 | 1 | 113417353 | 113468865 | rs749956 | 6 E-04 | 79 | 392 | 7 E-03 |
| ST20 | 15 | 77978236 | 78003132 | rs1814345 | 4 E-05 | 53 | 163 | 7 E-03 |
| HSPE1 | 2 | 198073364 | 198076416 | rs7585486 | 3 E-03 | 25 | 125 | 7 E-03 |
| IL15 | 4 | 142777203 | 142874062 | rs7677372 | 5 E-03 | 102 | 481 | 7 E-03 |
| TRIM56 | 7 | 100515505 | 100520609 | rs10231438 | 3 E-04 | 45 | 166 | 7 E-03 |
| OR2W1 | 6 | 29119968 | 29120931 | rs2143574 | 4 E-03 | 65 | 340 | 7 E-03 |
| FOXP1 | 3 | 71087425 | 71715830 | rs830628 | 2 E-04 | 314 | 777 | 7 E-03 |
| DLX1 | 2 | 172658453 | 172662647 | rs10930506 | 3 E-03 | 30 | 93 | 7 E-03 |
| MAD2L2 | 1 | 11657123 | 11674265 | rs7526268 | 4 E-03 | 17 | 64 | 7 E-03 |
| CCR8 | 3 | 39346218 | 39351077 | rs1113160 | 8 E-04 | 39 | 157 | 7 E-03 |
| ANGPTL5 | 11 | 101266614 | 101292463 | rs4576789 | 2 E-04 | 59 | 223 | 7 E-03 |
| GYPB | 4 | 145136706 | 145159946 | rs13103072 | 6 E-03 | 29 | 197 | 7 E-03 |
| C3orf58 | 3 | 145173602 | 145193895 | rs4527375 | 3 E-03 | 77 | 507 | 7 E-03 |
| DCUN1D5 | 11 | 102438030 | 102468079 | rs2509112 | 3 E-03 | 45 | 214 | 7 E-03 |

**Table S4c. Top Gene-Based Association Test Results from GWAS of Heterophily**

| Gene | Chr | Start | Stop | Best SNP | SNP p value | # SNPs | Test Stat | Gene p value |
|---|---|---|---|---|---|---|---|---|
| MKL1 | 22 | 39136237 | 39362636 | rs5757979 | 3 E-05 | 60 | 235 | 7 E-03 |
| C1orf9 | 1 | 170768882 | 170847596 | rs10911843 | 7 E-03 | 75 | 274 | 7 E-03 |
| ZSCAN23 | 6 | 28508410 | 28519258 | rs2531806 | 8 E-04 | 63 | 348 | 7 E-03 |
| LIPF | 10 | 90414073 | 90428552 | rs11202817 | 7 E-04 | 107 | 487 | 7 E-03 |
| DDHD1 | 14 | 52582490 | 52689750 | rs2254182 | 8 E-04 | 90 | 458 | 7 E-03 |
| FLJ21438 | 19 | 15423437 | 15436382 | rs887898 | 5 E-03 | 21 | 83 | 8 E-03 |
| TBC1D8 | 2 | 100990121 | 101134278 | rs7585873 | 2 E-03 | 145 | 427 | 8 E-03 |
| EXTL1 | 1 | 26220857 | 26235541 | rs12067677 | 1 E-03 | 70 | 312 | 8 E-03 |
| OR2B3P | 6 | 29162062 | 29163004 | rs2143574 | 4 E-03 | 63 | 311 | 8 E-03 |
| COQ10B | 2 | 198026475 | 198048096 | rs7585486 | 3 E-03 | 28 | 138 | 8 E-03 |
| BACH1 | 21 | 29593090 | 29656086 | rs2832296 | 2 E-03 | 43 | 255 | 8 E-03 |
| ALOX5AP | 13 | 30207668 | 30236556 | rs12429692 | 4 E-04 | 132 | 403 | 8 E-03 |
| ATXN7L2 | 1 | 109828083 | 109836879 | rs3738772 | 9 E-04 | 58 | 240 | 8 E-03 |
| FMO3 | 1 | 169326659 | 169353583 | rs10912478 | 3 E-03 | 71 | 289 | 8 E-03 |
| GBP1 | 1 | 89290574 | 89303631 | rs1329119 | 4 E-03 | 81 | 331 | 8 E-03 |
| TGS1 | 8 | 56848344 | 56900559 | rs11998723 | 4 E-03 | 81 | 389 | 8 E-03 |
| C9orf93 | 9 | 15543096 | 15961897 | rs2039980 | 1 E-03 | 357 | 1749 | 8 E-03 |
| LPIN3 | 20 | 39402973 | 39422636 | rs17181845 | 3 E-03 | 52 | 196 | 8 E-03 |
| KLRA1 | 12 | 10632343 | 10643701 | rs6488305 | 6 E-04 | 143 | 516 | 8 E-03 |
| FASLG | 1 | 170894807 | 170902635 | rs2859228 | 7 E-03 | 36 | 122 | 8 E-03 |
| WDR4 | 21 | 43136272 | 43172747 | rs6586252 | 4 E-03 | 81 | 375 | 8 E-03 |
| HDAC9 | 7 | 18501893 | 19003517 | rs6959028 | 3 E-04 | 383 | 922 | 8 E-03 |
| CYB561D1 | 1 | 109838244 | 109843000 | rs3738772 | 9 E-04 | 57 | 228 | 8 E-03 |
| TCEB1 | 8 | 75021187 | 75046900 | rs12547746 | 2 E-03 | 59 | 254 | 8 E-03 |
| AMIGO1 | 1 | 109850968 | 109853859 | rs3738772 | 9 E-04 | 58 | 226 | 8 E-03 |
| NEK11 | 3 | 132228416 | 132551993 | rs12639350 | 1 E-03 | 104 | 471 | 8 E-03 |
| ZNF311 | 6 | 29070572 | 29081016 | rs2143574 | 4 E-03 | 57 | 287 | 9 E-03 |
| DOPEY2 | 21 | 36458708 | 36588442 | rs2242810 | 1 E-03 | 122 | 336 | 9 E-03 |
| CCDC101 | 16 | 28472757 | 28510610 | rs4788073 | 4 E-03 | 22 | 107 | 9 E-03 |
| GBP2 | 1 | 89345897 | 89364387 | rs1329119 | 4 E-03 | 67 | 328 | 9 E-03 |
| CBR3 | 21 | 36429132 | 36440730 | rs2242810 | 1 E-03 | 60 | 192 | 9 E-03 |
| EGFLAM | 5 | 38294289 | 38501338 | rs2589811 | 3 E-05 | 209 | 580 | 9 E-03 |
| C20orf71 | 20 | 31268795 | 31279220 | rs6141900 | 5 E-03 | 40 | 205 | 9 E-03 |
| ADNP2 | 18 | 75967902 | 75999219 | rs4402630 | 2 E-04 | 90 | 314 | 9 E-03 |
| SENP7 | 3 | 102525807 | 102714775 | rs2162303 | 4 E-03 | 145 | 755 | 9 E-03 |
| OTOL1 | 3 | 162697289 | 162704424 | rs6797876 | 6 E-06 | 81 | 448 | 9 E-03 |
| OR2V2 | 5 | 180514548 | 180515496 | rs17705959 | 1 E-03 | 20 | 70 | 9 E-03 |
| LDHAL6A | 11 | 18434006 | 18457723 | rs10500836 | 2 E-03 | 45 | 169 | 9 E-03 |
| OR4L1 | 14 | 19598043 | 19598982 | rs1959641 | 2 E-03 | 88 | 358 | 9 E-03 |
| TMEM68 | 8 | 56813873 | 56848439 | rs11998723 | 4 E-03 | 68 | 336 | 9 E-03 |

| Correlation Type | homophily | homophily | homophily | heterophily | heterophily | heterophily |
| --- | --- | --- | --- | --- | --- | --- |
| Threshold | Top 1% | Top 5% | Top 25% | Top 1% | Top 5% | Top 25% |
| Genes in the Test Set | 172 | 499 | 4352 | 173 | 870 | 4352 |
| Genes in the Reference Set | 17412 | 17412 | 17412 | 17412 | 17412 | 17412 |
| Test Set genes linked to at least one term in the GOSlim database | 147 | 407 | 3582 | 144 | 733 | 3598 |
| Reference set genes linked to at least one term in the GOSlim database | 14276 | 14276 | 14276 | 14276 | 14276 | 14276 |
| Test Set genes linked to at least one term in the KEGG database | 62 | 164 | 1326 | 55 | 275 | 1369 |
| Reference set genes linked to at least one term in the KEGG database | 5322 | 5322 | 5322 | 5322 | 5322 | 5322 |

**Table S5. Sample sizes for KEGG and GOSlim gene set analysis** before and after pruning genes from the test set and reference set that can be linked to at least one term in the database.

| Quintile | Range of SNP Correlations | Mean SNP Correlation | Composite of Multiple Signals (CMS) Score | | | |
|---|---|---|---|---|---|---|
| | | | Friends | | Strangers | |
| | | | Mean | SE | Mean | SE |
| 1 | < –0.03 | –0.05 | –8.044 | 0.025 | –7.995 | 0.027 |
| 2 | –0.03 to –0.01 | –0.02 | –8.034 | 0.022 | –8.012 | 0.022 |
| 3 | –0.01 to 0.01 | 0.00 | –8.015 | 0.022 | –8.037 | 0.022 |
| 4 | 0.01 to 0.03 | 0.02 | –8.023 | 0.022 | –8.009 | 0.022 |
| 5 | > 0.03 | 0.05 | –7.957 | 0.026 | –8.018 | 0.024 |

**Table S6. Mean Composite of Multiple Signals (CMS) score, by SNP correlation quintile.** Shown in Fig.3 of the main text. Taken from a weighted least squares (WLS) regression of CMS on fixed effects for each SNP correlation quintile for 1,468,013 SNPs (weights are inverse squared standard errors for each of the GWAS correlation coefficients). Newey-West standard errors correct for linkage disequilibrium and serial correlation in CMS and GWAS coefficients.

|  | Coefficient | Standard Error | t Statistic | p Value |
|---|---|---|---|---|
| **SNP is in Top Quintile of most Homophilic SNPs** | **0.07** | **0.02** | **2.95** | **0.003** |
| *Constant* | –8.03 | 0.02 | –489.73 | 0.000 |

**Table S7. Positively Correlated SNPs between *friends* are more likely to be under positive selection.** Weighted Least Squares (WLS) regression of CMS score on an indicator variable for homophily (weights are inverse squared standard errors for each of the GWAS correlation coefficients). $N$ = 1,468,013 SNPs. Newey-West standard errors correct for linkage disequilibrium and serial correlation in CMS and GWAS coefficients. Newey-West adjusted model fit $F(2,1468012) = 8.72$, $p = 0.003$.

|  | Coefficient | Standard Error | t Statistic | p Value |
|---|---|---|---|---|
| *SNP is in Top Quintile of most Homophilic SNPs* | –0.00 | 0.02 | –0.45 | 0.86 |
| *Constant* | –8.01 | 0.02 | –471.92 | 0.00 |

**Table S8. Correlation in SNPs between *strangers* is not related to positive selection.** Weighted Least Squares (WLS) regression of CMS score on an indicator variable for homophily (weights are inverse squared standard errors for each of the GWAS correlation coefficients). $N$ = 1,468,013 SNPs. Newey-West standard errors correct for linkage disequilibrium and serial correlation in CMS and GWAS coefficients. Newey-West adjusted model fit $F\,(2,1468012) = 0.03$, $p = 0.86$.

|  | Coefficient | Standard Error | t Statistic | p Value |
| --- | --- | --- | --- | --- |
| **Homophily (Positive Correlation)** | **1.27** | **0.58** | **2.18** | **0.03** |
| *Heterophily (Negative Correlation)* | 0.26 | 0.54 | 0.48 | 0.63 |
| *Constant* | –8.03 | 0.02 | –396.11 | 0.00 |

**Table S9. Positively Correlated SNPs between *friends* are more likely to be under positive selection.** Weighted Least Squares (WLS) regression of CMS score on measures of homophily and heterophily (weights are inverse squared standard errors for each of the GWAS correlation coefficients). $N$ = 1,468,013 SNPs. Newey-West standard errors correct for linkage disequilibrium and serial correlation in CMS and GWAS coefficients. Newey-West adjusted model fit $F(2,1468012) = 3.23$, $p = 0.04$.

|  | Coefficient | Standard Error | t Statistic | p Value |
|---|---|---|---|---|
| *Homophily (Positive Correlation)* | 0.17 | 0.57 | 0.31 | 0.77 |
| *Heterophily (Negative Correlation)* | −0.61 | 0.57 | −1.07 | 0.28 |
| *Constant* | −8.03 | 0.02 | −394.23 | 0.00 |

**Table S10. Correlation in SNPs between *strangers* is not related to positive selection.** Weighted Least Squares (WLS) regression of CMS score on measures of homophily and heterophily (weights are inverse squared standard errors for each of the GWAS correlation coefficients). $N$ = 1,468,013 SNPs. Newey-West standard errors correct for linkage disequilibrium and serial correlation in CMS and GWAS coefficients. Newey-West adjusted model fit $F$ (2,1468012) = 0.57, $p$ = 0.56.